\documentclass[twocolumn,aps,prx,amsmath,amssymb,superscriptaddress]{revtex4}
\usepackage{graphicx,dcolumn,bm,color,framed}
\usepackage{ulem}
\usepackage{bm}

\begin{document}

\title{The role of electron-electron collisions for charge and heat transport at intermediate temperatures}

\author{Woo-Ram Lee}
\affiliation{Department of Physics and Astronomy, The University of Alabama, Tuscaloosa, Alabama 35487, USA}
\affiliation{Center for Materials for Information Technology, The University of Alabama, Tuscaloosa, Alabama 35401, USA}
\author{Alexander M. Finkel'stein}
\affiliation{Department of Physics and Astronomy, Texas A\&M University, College Station, Texas 77843-4242, USA}
\affiliation{Department of Condensed Matter Physics, The Weizmann Institute of Science, Rehovot 76100, Israel}
\author{Karen Michaeli}
\affiliation{Department of Condensed Matter Physics, The Weizmann Institute of Science, Rehovot 76100, Israel}
\author{Georg Schwiete}
\affiliation{Department of Physics and Astronomy, The University of Alabama, Tuscaloosa, Alabama 35487, USA}
\affiliation{Center for Materials for Information Technology, The University of Alabama, Tuscaloosa, Alabama 35401, USA}


\begin{abstract}
Electric, thermal and thermoelectric transport in correlated electron systems probe different aspects of the many-body dynamics, and thus provide complementary information. These are well studied in the low- and high-temperature limits, while the experimentally important intermediate regime, in which elastic and inelastic scattering are both important, is less understood. To fill this gap, we provide comprehensive solutions of the Boltzmann equation in the presence of an electric field and a temperature gradient for two different cases: First, when electron-electron collisions are treated within the relaxation-time approximation while the full momentum dependence of electron-impurity scattering is included and, second, when the electron-impurity scattering is momentum-independent, but the electron-electron collisions give rise to a momentum-dependent inelastic scattering rate of the Fermi-liquid type. We find that for Fermi-liquid as well as for Coulomb interactions, both methods give the same results for the leading temperature dependence of the transport coefficients. Moreover, the inelastic relaxation rate enters the electric conductivity and the Seebeck coefficient only when the momentum dependence of the electron-impurity collisions, analytical or non-analytical, is included. Specifically, we show that inelastic processes only mildly affect the electric conductivity, but can generate a non-monotonic dependence of the Seebeck coefficient on temperature and even a change of sign. Thermal conductivity, by contrast, always depends on the inelastic scattering rate even for a constant elastic relaxation rate. 

\end{abstract}

\maketitle

\section{Introduction}

The temperature dependence of transport coefficients in an electron liquid is intimately related to the effectiveness of various microscopic scattering mechanisms \cite{Ziman01,Gantmakher12,Pal12}. At low temperatures, inelastic electron-electron scattering is usually not important, and the currents relax through elastic scattering of electrons by impurities. Electron-electron scattering plays a more significant role at higher temperatures. Once it dominates over electron-impurity scattering, the system approaches the hydrodynamic limit~\cite{Andreev11,Mahajan13,Narozhny17,Guo17,Levchenko17,Lucas18a,Hartnoll18}, and current relaxation occurs collectively~\cite{Remark_phonons}. These two limiting cases have been extensively studied \cite{Castellani84,Altshuler85,Castellani87,finkelstein90,Fabrizio91,Livanov91,Zala01,Raimondi04,Catelani05,Niven05,Michaeli09,Shastry09,Andreev11,Mahajan13,Schwiete14a,Schwiete16a,Schwiete16b,Narozhny17,Guo17,Levchenko17,Lucas18a,Hartnoll18}. By contrast, the intermediate-temperature regime, where electron-electron scattering and electron-impurity scattering are comparable, is less understood \cite{Principi15,Lucas18,Remark_graphene}. We expect the transport properties at intermediate temperatures to be highly sensitive to the nature of both elastic and inelastic collisions. Since many experimental measurements of electric, thermal, and thermoelectric transport coefficients may well be falling in this regime, our goal here is to close this gap.

To illustrate the importance of elastic and inelastic scattering processes for transport, let us consider the electric conductivity $\sigma$: In the absence of inelastic collisions, each electron loses its momentum only when it is directly scattered by impurities. The electric current is carried by many independent channels, each corresponding to a single-particle state. Consequently, the electric conductivity is proportional to the weighted sum of single-particle momentum relaxation times $\sigma\propto\sum_{\mathbf{p}} \tau_{ei,\mathbf{p}}{\bf p}^2 \partial n_F(\xi_\mathbf{p})/\partial\xi_{\mathbf{p}}$, where the relaxation time $\tau_{ei,\mathbf{p}}$ may depend on the electron momentum $\mathbf{p}$, and $n_F(\xi_{\mathbf{p}})$ is Fermi-Dirac distribution at energy $\xi_{\mathbf{p}}$. In this limit, the conductivity is dominated by the most conducting channel, as for resistors connected in parallel. This result, known as the Drude formula,  does not hold in the hydrodynamic limit where the \textit{total momentum} of the electron fluid relaxes by impurities only after inelastic interactions give rise to  a uniform flow. As a consequence of the collective nature of the flow in this regime, the strongest elastic relaxation process determines the conductivity. Hence, the system is equivalent to many resistors connected in series, and $\sigma\propto1/(\sum_{\mathbf{p}}\tau_{ei,\mathbf{p}}^{-1}{\bf p}^2 \partial n_F(\xi_{\mathbf{p}})/\partial\xi_{\mathbf{p}})$. Importantly, the conductivity in either limit is independent of the details of the inelastic scattering.

In the intermediate regime, it has been shown~\cite{Keyes58,Gantmakher12} that within the relaxation-time approximation (RTA) for the electron-electron scattering $\tau_{ee}$, both scattering times, $\tau_{ei,\mathbf{p}}$ and  $\tau_{ee}$, can enter the electric conductivity. Importantly, $\tau_{ee}$ affects the conductivity only when the single-particle momentum relaxation by impurities depends on the energy $\xi_{\bf p}$~\cite{Remark_Fermi_surface}. If $\tau_{ei,\mathbf{p}}\sim\text{const}$, the electric conductivity is described by the Drude formula regardless of the strength of inelastic scattering. The relatively weak sensitivity of the electric conductivity to electron-electron collisions is a consequence of the inability of these collisions to relax the total momentum of the electron liquid. By contrast, inelastic scattering alone is typically sufficient to generate a finite thermal conductivity. This is because inelastic collisions relax the thermal current---loosely the energy-weighted sum of single particle momenta---despite conserving the total energy and momentum.

The RTA leads to a huge simplification of the full kinetic problem. In this approximation, all modes that are not protected by conservation laws relax towards equilibrium with the same temperature-dependent rate. It is important to remember, however, that a generic electron-electron collision integral gives rise to a scattering rate that depends not only on temperature but also on the energy of the scattered particle. One important goal of this work is to explore the sensitivity of the various transport coefficients to the exact form of the inelastic collisions.

Here, we are interested in various transport coefficients studied in a framework of a semi-phenomenological model where the collision integral is described by the sum of two scattering processes, arising from: (i) electron-electron interactions and (ii) scattering of electrons by impurities.
Let us emphasize that elastic scattering may dramatically modify the inelastic scattering rate. Also, electron-electron interactions modify the elastic scattering rate. We assume, however, that the various renormalizations have already been incorporated, and our
goal is to understand how the two terms in the collision integral interfere with each other when calculating transport coefficients.

The main focus of this work lies on the effect that momentum-dependent scattering processes, both elastic and inelastic ones, have on various transport properties. Solving this problem even within the simplified model of two independent contributions to the collision integral is a formidable task. Fortunately, an exact solution can be found in the special case where  the momentum dependence of one scattering process is included within the Boltzmann equation,  while the other is kept  constant. The main result of our work is to derive response functions  of electrons in the presence of an electric field and a temperature gradient in the framework of the Boltzmann equation with two collision terms.

We start with an extension of the self-consistent solution of Keyes~\cite{Keyes58} to the thermal and thermoelectric transport coefficients, in addition to the electric conductivity. As anticipated, we find that the inelastic scattering time affects all transport coefficients. The magnitude of the interaction effects is, however, qualitatively different for each of them. In particular, we show that thermal conductivity strongly depends on the inelastic scattering time, which affects this transport coefficient even in the presence of a momentum-independent elastic relaxation rate. By contrast, thermoelectric transport coefficients are independent of  the inelastic scattering time when $\tau_{ei,\mathbf{p}}$ is constant, similar to the electric conductivity. If the elastic scattering rate is momentum-dependent, however, it induces a $\tau_{ee}$ dependence of the Seebeck coefficient. Consequently, this transport coefficient may develop a characteristic temperature dependence, including the possibility of non-monotonic behavior and a change of sign. All obtained results are applicable for both analytic and non-analytic energy dependences of the relaxation rate.

The energy dependence of the electron-impurity scattering rate can have a smooth analytic component in the vicinity of the Fermi energy. This component may originate, for example, from a non-constant density of states or through the momentum dependence of the scattering amplitude for the disorder potential. The energy dependence may also have non-analytic contributions. This occurs when Friedel oscillations modify the scattering of electrons by impurities. Such an effect has been shown to generate a linear in $T$ correction to the electric conductivity of two-dimensional electron liquids~\cite{Gold86,DasSarma99,Zala01,Gornyi04,Gold11} in the ballistic regime and under the assumption that $1/\tau_{ee}\ll 1/\tau_{ei}$. As we show here, the non-analytic form of the elastic scattering rate can induce a dependence of transport coefficients on the inelastic scattering rate. Our results suggest that the existing theory for the transport coefficients needs to be generalized to include inelastic scattering at the intermediate temperatures of main interest in this paper ($1/\tau_{ee}\approx 1/\tau_{ei}$).

We compare our result for the thermal conductivity and the solution in the presence of momentum-dependent electron-electron scattering rate and a constant elastic scattering rate. For this purpose, we apply the method introduced in Ref.~\cite{Bennett69} for calculating the thermal conductivity of impure three-dimensional Fermi liquids (FLs). Based on the strong $\tau_{ee}$ dependence of the thermal conductivity found within the RTA, we expected the result to significantly change for the different type of interactions. We find, however, that the thermal conductivities obtained {with the method described in Ref.~\cite{Bennett69} can be matched to the expression found with the generalized Keyes method by properly fixing a single parameter, $\tau_{ee}$. This surprising observation along  with the exact expression for the conductivities derived within the RTA for inelastic scattering is the main result of our work. Furthermore, our calculation for a specific sample interaction provides important insight into what type of interactions can be described using the RTA and when strong deviations are expected.

The remainder of the paper is organized as follows: In Sec.~\ref{Sec:Boltzmann}, we introduce the Boltzmann equation in the presence of electric and thermal driving forces.  In Sec.~\ref{Sec:Keyes approach}, we generalize Keyes approach, and find the non-equilibrium distribution function within the RTA. We use this distribution function to compute the electric, thermal and thermoelectric transport coefficients. The eigenfunction expansion of Ref.~\cite{Bennett69} is employed in Sec.~\ref{Sec:e-e collision integral beyond RTA} to derive the conductivities in the presence of FL and Coulomb interactions. In Sec.~\ref{Sec:Discussion:Momentum-independent}, we then compare the results obtained with both methods for momentum-independent elastic scattering. In Sec.~\ref{Sec:Discussion:Momentum-dependent}, we analyze the combined effect of electron-impurity and electron-electron scattering when the elastic scattering rate depends on momentum. The Seebeck coefficient and its unique temperature dependence due to inelastic collisions are discussed in Sec.~\ref{Sec:Seebeck}. Final remarks appear in the conclusions, Sec.~\ref{Sec:Conclusion}. Several appendices cover technical aspects of the discussion provided in the main text.

\section{Linearized Boltzmann equation}
\label{Sec:Boltzmann}

We study electric, thermal, and thermoelectric transport in an electron system using the Boltzmann equation \cite{Ziman01} with a collision integral that describes two scattering processes: $I\{f\} = I_{ei}\{f\} + I_{ee}\{f\}$, where $I_{ei}$ ($I_{ee}$) stands for electron-impurity (electron-electron) scattering, and $f$ is the distribution function. For the sake of simplicity, we describe $I_{ei}$ in the RTA throughout this paper. In this approximation, the distribution function $f({\bf r},{\bf p},t)$ relaxes towards the angularly averaged distribution $\langle f({\bf r},{\bf p},t)\rangle$, and we formulate
\begin{align}
I_{ei}\{f\}
= - \frac{f({\bf r},{\bf p},t) - \langle f({\bf r},{\bf p},t) \rangle}{\tau_{ei,{\bf p}}}.
\label{Coll_ei}
\end{align}
Here, we include a momentum dependence in the characteristic time $\tau_{ei,{\bf p}}$. We do not distinguish here between the single particle scattering time due to disorder, $\tau_{ei,{\bf p}}$, and the transport scattering time. Generalizing the result to cases in which these two times are unequal is straightforward \cite{Remarktautr}.

To find the transport coefficients for steady-state currents in linear response, the distribution function is expanded as $f({\bf r}, {\bf p}) \approx n_{F}(\xi_{\bf p}) + \delta f_{\bf p}$. Here, $n_{F}(\xi_{\bf p}) = [\exp(\beta\xi_{\bf p}) + 1]^{-1}$ is the Fermi-Dirac distribution with $\beta = (k_BT)^{-1}$, and $\xi_{\bf p} = \epsilon_{\bf p} - \mu$ with the chemical potential $\mu$. The small deviation from equilibrium, $\delta f_{\bf p}$, is induced by an electric field ${\bf E}$ or a temperature gradient $\nabla_{\bf r} T$, which will be treated in linear response only. In Eq.~\eqref{Coll_ei}, the angularly averaged distribution function may be approximated by the equilibrium distribution $n_{F}$, since $\langle\delta f_{\bf p}\rangle=0$ in linear response. Consequently, the linearized Boltzmann equation reads
\begin{align}
\left( -e{\bf E} - \xi_{\bf p} \frac{\nabla_{\bf r}T}{T} \right) \cdot \boldsymbol{v}_{\bf p} \frac{\partial n_{F}(\xi_{\bf p})}{\partial\xi_{\bf p}}
= - \frac{\delta f_{\bf p}}{\tau_{ei,{\bf p}}} + I_{ee}\{\delta f\}.
\label{BoltzmannEq_Linearized}
\end{align}
Here, the collision integral $I_{ee}$ is also linearized.

In the following sections, we will discuss solutions to Eq.~\eqref{BoltzmannEq_Linearized} for different types of electron-electron collision integrals. For simplicity, we assume a quadratic band dispersion $\epsilon_{\bf p}=p^2/2m$.

\section{Relaxation-time approximation for $I_{ee}$}
\label{Sec:Keyes approach}

It is instructive to start with a simple theory in which both collision integrals, $I_{ei}$ and $I_{ee}$, are described in the RTA. A model of this type was studied by Keyes for the electric conductivity~\cite{Keyes58}. We generalize this approach, and derive closed-form solutions for the thermal conductivity and the Seebeck coefficient. Our results provide an instructive example of a non-perturbative description of the linear conductivities in the entire range between the impurity-dominated and the interaction-dominated transport regimes.

The electron-electron collision integral $I_{ee}$ in the RTA differs from $I_{ei}$ in an important respect: Electron-impurity collisions change the momenta of the scattered electrons. As a consequence, in linear response, $I_{ei}$ describes a relaxation towards the Fermi-Dirac distribution in the laboratory frame in which impurities are at rest. In contrast, electron-electron collisions conserve the total momentum of the participating particles. The collision integral $I_{ee}$ therefore describes a relaxation towards a ``drifting" Fermi-Dirac distribution function $n_F^{({\rm cm})}$,
\begin{align}
I_{ee}\{f\}
= - \frac{f({\bf r},{\bf p}) - n_F^{({\rm cm})}(\bf p)}{\tau_{ee}},
\label{Coll_ee}
\end{align}
where $\tau_{ee}$ is the characteristic time for electron-electron scattering and cm stands for ``center of mass". The ``drifting" Fermi-Dirac distribution function is related to the Fermi-Dirac distribution in the laboratory frame as follows,
\begin{align}
n_F^{({\rm cm})}({\bf p}) = n_F(\epsilon_{\bf p}-\mu-{\boldsymbol v}_{\rm cm}\cdot{\bf p}).\label{eq:ncm}
\end{align}
The drift velocity ${\boldsymbol v}_{\rm cm}$ appears here because the system is kept in a non-equilibrium steady state; this quantity can be interpreted as the velocity of the center-of-mass motion of electrons for the case of the quadratic dispersion that we consider. It is worth mentioning that, unlike the case of electron-impurity scattering, the form  of $I_{ee}$ given above in the RTA is very constrained: $\tau_{ee}$ must be independent of momentum for Eq.~\eqref{Coll_ee} to be consistent with all conservation laws, the conservation of particle number and energy besides the already mentioned momentum conservation.

Note that the drift velocity ${\boldsymbol v}_{\rm cm}\{f\}$ itself depends on the non-equilibrium distribution. This is because the total momentum associated with the distribution functions $f$ and $n_F^{({\rm cm})}$ must be equal. Moreover, the inverse temperature $\beta\{f\}$ [implicit in Eq.~\eqref{eq:ncm}], as well as the chemical potential $\mu\{f\}$,
also depend on $f$. For these reasons, linearizing the collision integral $I_{ee}$ is not entirely straightforward. However, as we explain in more detail in Appendix~\ref{Appendix:RTAforSee}, for the purpose of our calculation, we may take $\mu$ and $\beta$ as constant. In the next subsection, we obtain the connection of ${\boldsymbol v}_{\rm cm}\{f\}$ with the electric field and the temperature gradient.

In our problem, a finite ${\boldsymbol v}_{\rm cm}$ only exists due to the driving of the system by either the electric field or the temperature gradient. This is why ${\boldsymbol v}_{\rm cm}$ is a small quantity in the linear response regime, and we may linearize $n_F^{({\rm cm})}({\bf p}) \approx (1 - {\boldsymbol v}_{\rm cm} \cdot {\bf p} \partial_{\xi_{\bf p}}) n_F(\xi_{\bf p})$. Using this simplification, one obtains
\begin{align}
I_{ee}\{f\}=-\frac{\delta f_{\bf p} + {\boldsymbol v}_{\rm cm}\cdot {\bf p}\partial n_F(\xi_{\bf p})/\partial\xi_{\bf p}}{\tau_{ee}}.
\label{Seelinear}
\end{align}
Here, the center-of-mass velocity can be expressed as ${\boldsymbol v}_{\rm cm} = s \int_{\bf p} {\bf p}\delta f_{\bf p} / (\mathcal{N}m)$ with the particle density $\mathcal{N} = s\int_{\bf p} n_F(\xi_{\bf p})$ and the spin degeneracy $s=2$.

In this paper, we use the short notation for integrals $\int_{\bf p} = \int d{\bf p}/(2\pi)^d$ as well as units with $\hbar=c=k_B=1$.

\subsection{Non-equilibrium distribution}
\label{Subsec:NED}

Solving the linearized Boltzmann equation Eq.~\eqref{BoltzmannEq_Linearized} with the collision integral $I_{ee}$ given in Eq.~\eqref{Seelinear} is still a formidable task. This is because the explicit dependence on $\delta f_{\bf p}$ is accompanied by the implicit dependence through ${\boldsymbol v}_{\rm cm}$. Resolving the explicit dependence first, one finds
\begin{align}
\delta f_{\bf p}
= \tilde{\tau}_{\bf p} \boldsymbol{v}_{\bf p} \cdot \left( e \tilde{{\bf E}} + \xi_{\bf p} \frac{\nabla_{\bf r} T}{T} \right) \frac{\partial n_F(\xi_{\bf p})}{\partial \xi_{\bf p}},
\label{TrialSolution}
\end{align}
with the effective electric field 
\begin{align}
\tilde{{\bf E}} = {\bf E} - \frac{m{\boldsymbol v}_{\rm cm}}{e\tau_{ee}},
\label{EffectiveElectricField}
\end{align}
and the total scattering rate
\begin{align}
\frac{1}{\tilde{\tau}_{\bf p}}
= \frac{1}{\tau_{ei,{\bf p}}} + \frac{1}{\tau_{ee}}.
\label{TotalScatteringRate}
\end{align}
The total scattering rate satisfies the Matthiessen's rule, which is a direct consequence of presenting the collision term as a sum of two terms, $I = I_{ei} + I_{ee}$.

The key observation of Ref.~\cite{Keyes58} is that the drift velocity may be found by computing ${\boldsymbol v}_{\rm cm} = s \int_{\bf p} {\bf p}\delta f_{\bf p} / (\mathcal{N}m)$ with the help of Eq.~\eqref{TrialSolution}. This results in a \textit{self-consistent} equation for ${\boldsymbol v}_{\rm cm}$ that is easily solved. After some algebra, we arrive at
\begin{align}
m{\boldsymbol v}_{\rm cm}
= \frac{\tau_{ee}}{\tau_{ee} - \langle\!\langle \tilde{\tau}_{\bf p} \rangle\!\rangle} \bigg( - \langle\!\langle \tilde{\tau}_{\bf p} \rangle\!\rangle e{\bf E} - \langle\!\langle \xi_{\bf p} \tilde{\tau}_{\bf p} \rangle\!\rangle \frac{\nabla_{\bf r} T}{T} \bigg).
\label{ElectricFieldCorrection}
\end{align}
Here, $\left\langle\!\left\langle\dots\right\rangle\!\right\rangle$ denotes the average
\begin{align}
\langle\!\langle X_{\bf p} \rangle\!\rangle
= - \frac{2s}{d\mathcal{N}} \int_{\bf p} X_{\bf p} (\xi_{\bf p} + \mu) \frac{\partial n_F(\xi_{\bf p})}{\partial \xi_{\bf p}},
\label{FluctuationAverage}
\end{align}
with normalization $\langle\!\langle 1\rangle\!\rangle=1$, for which the dimensionality $d$ and the factor $\xi_{\bf p}+\mu$ appear due to the angular averaging, $\boldsymbol{v}_{\bf p} (\boldsymbol{v}_{\bf p} \cdot {\bf Y}) \rightarrow v_{\bf p}^2 {\bf Y}/d$ for ${\bf Y} \in \{{\bf E}, \nabla_{\bf r}T\}$. By inserting Eq.~\eqref{ElectricFieldCorrection} into Eqs.~\eqref{TrialSolution}-\eqref{TotalScatteringRate}, we obtain the deviation from the equilibrium distribution function in response to the electric field and the temperature gradient, respectively, 
\begin{align}
\delta f_{\bf p}^E
& = \frac{\tau_{ee} \tilde{\tau}_{\bf p}}{\tau_{ee} - \langle\!\langle \tilde{\tau}_{\bf p} \rangle\!\rangle} \boldsymbol{v}_{\bf p} \cdot e{\bf E} \frac{\partial n_F(\xi_{\bf p})}{\partial \xi_{\bf p}},
\label{DistFluc_E}
\\
\delta f_{\bf p}^T
& = \tilde{\tau}_{\bf p} \bigg( \xi_{\bf p} + \frac{\langle\!\langle \xi_{\bf p} \tilde{\tau}_{\bf p} \rangle\!\rangle}{\tau_{ee} - \langle\!\langle \tilde{\tau}_{\bf p} \rangle\!\rangle} \bigg) \boldsymbol{v}_{\bf p} \cdot \frac{\nabla_{\bf r} T}{T} \frac{\partial n_F(\xi_{\bf p})}{\partial \xi_{\bf p}}.
\label{DistFluc_T}
\end{align}
Note that in metallic systems $\langle\!\langle \xi_{\bf p} \tilde{\tau}_{\bf p} \rangle\!\rangle$ tends to be small at low temperatures (in view of $\langle\!\langle \xi_{\bf p} \rangle\!\rangle\propto T^2/\epsilon_F$). We have to keep it for two reasons: (i) We are interested in a wide range of temperatures, and (ii) thermoelectricity is determined by $\langle\!\langle \xi_{\bf p} \tilde{\tau}_{\bf p} \rangle\!\rangle$ and $\langle\!\langle \xi_{\bf p}\rangle\!\rangle$.

It is instructive to first study the case of constant $\tau_{ei}$, for which a drastic simplification occurs,
\begin{align}
m{\boldsymbol v}_{\rm cm}
= \tau_{ei} \bigg( - e{\bf E} - \langle\!\langle \xi_{\bf p} \rangle\!\rangle \frac{\nabla_{\bf r} T}{T} \bigg),
\label{ElectricFieldCorrection_ConstScatt}
\end{align}
and therefore,
\begin{align}
\delta f_{\bf p}^E
& = \tau_{ei}\boldsymbol{v}_{\bf p} \cdot e{\bf E}  \frac{\partial n_F(\xi_{\bf p})}{\partial \xi_{\bf p}},
\label{DistFluc_E_ConstScatt}
\\
\delta f_{\bf p}^T
& = \bigg( \tau_{ei} \langle\!\langle \xi_{\bf p}\rangle\!\rangle + \frac{\xi_{\bf p} - \langle\!\langle \xi_{\bf p}\rangle\!\rangle}{\tau_{ei}^{-1} + \tau_{ee}^{-1}} \bigg) \boldsymbol{v}_{\bf p} \cdot \frac{\nabla_{\bf r} T}{T} \frac{\partial n_F(\xi_{\bf p})}{\partial \xi_{\bf p}}.
\label{DistFluc_T_ConstScatt}
\end{align}
We immediately notice that $\delta f_{\bf p}^E$ depends only on the constant electron-impurity scattering time $\tau_{ei}$, and is independent of the electron-electron collisions. The underlying reason is that $\delta f_{\bf p}^E$ is a zero mode of the collision integral $I_{ee}$. The entire Fermi surface is shifted by $\delta {\bf p} = m\boldsymbol{v}_{\rm cm}^E$ with the drift velocity $\boldsymbol{v}_{\rm cm}^E = -e{\bf E}\tau_{ei}/m$. This argument remains valid beyond the RTA, because the relevant zero mode of $I_{ee}$ is a result of momentum conservation during electron-electron collisions. For a momentum-dependent electron-impurity scattering time $\tau_{ei,{\bf p}}$, electron-electron collisions affect the response of the electron gas to the electric field. The inelastic collisions do not change the current directly. Rather, they modify the occupation of states in different energy shells that determine the strength of scattering by impurities.

The situation is fundamentally different for thermal driving, because the corresponding force is proportional to $\xi_{\bf p}$, and the momentum dependence is inherent in this case. Therefore, the second term in Eq.~\eqref{DistFluc_T_ConstScatt} survives in contrast to Eq.~\eqref{DistFluc_E_ConstScatt}, and $\delta f_{\bf p}^T$ depends on the electron-electron collisions even for constant electron-impurity scattering time $\tau_{ei}$. We notice that the Matthiessen's rule (dependence on $\tau_{ei}^{-1} + \tau_{ee}^{-1}$) works for the two scattering rates if $\langle\!\langle \xi_{\bf p}\rangle\!\rangle$ may be neglected.

\subsection{Transport coefficients}
\label{Subsec:conductivities}

The transport coefficients are fully determined by the non-equilibrium part of the distribution function. To find them, we insert Eqs.~\eqref{DistFluc_E}-\eqref{DistFluc_T} into the expressions for the electric and thermal current densities 
\begin{align}
\left(
\begin{array}{c}
{\bf J}_E
\\
{\bf J}_T
\end{array}
\right)
= s\int_{\bf p}
\left(
\begin{array}{c}
-e
\\
\xi_{\bf p}
\end{array}
\right)
\boldsymbol{v}_{\bf p} \delta f_{\bf p}
= \left(
\begin{array}{cc}
L_{EE} & L_{ET}
\\
L_{TE} & L_{TT}
\end{array}
\right) \left(
\begin{array}{c}
{\bf E}
\\
-\nabla_{\bf r} T
\end{array}
\right),
\end{align}
where $s=2$ is due to spin degeneracy. As a result, the electric conductivity reads
\begin{align}
\sigma
\equiv L_{EE}
= \frac{\mathcal{N}e^2\tau_{ee}}{m} \frac{\langle\!\langle \tilde{\tau}_{\bf p} \rangle\!\rangle}{\tau_{ee} - \langle\!\langle \tilde{\tau}_{\bf p} \rangle\!\rangle},
\label{Sigma_EE}
\end{align}
and the thermoelectric conductivities satisfy the Onsager reciprocal relation
\begin{align}
L_{ET}
& = \frac{L_{TE}}{T}
= - \frac{\mathcal{N}e\tau_{ee}}{mT} \frac{\langle\!\langle \xi_{\bf p} \tilde{\tau}_{\bf p} \rangle\!\rangle}{\tau_{ee} - \langle\!\langle \tilde{\tau}_{\bf p} \rangle\!\rangle}.
\label{Sigma_ET}
\end{align}
The Seebeck coefficient $S$, which measures the Seebeck effect, can be derived from Eqs.~\eqref{Sigma_EE} and \eqref{Sigma_ET}
\begin{align}
S \equiv \frac{L_{ET}}{L_{EE}} 
= - \frac{1}{eT} \frac{\langle\!\langle \xi_{\bf p} \tilde{\tau}_{\bf p} \rangle\!\rangle}{\langle\!\langle \tilde{\tau}_{\bf p} \rangle\!\rangle}.
\label{ThermoelectricPower}
\end{align}

The thermal conductivity is originally of the form
\begin{align}
L_{TT}
& = \frac{\mathcal{N}}{mT} \bigg( \langle\!\langle \xi_{\bf p}^2 \tilde{\tau}_{\bf p} \rangle\!\rangle + \frac{\langle\!\langle \xi_{\bf p} \tilde{\tau}_{\bf p} \rangle\!\rangle^2}{\tau_{ee} - \langle\!\langle \tilde{\tau}_{\bf p} \rangle\!\rangle} \bigg).
\label{Sigma_TT}
\end{align}
We note that all the response functions found above diverge for $\tau_{ei}\rightarrow \infty$. The divergence of $L_{TT}$ is specific to the thermal conductivity of open systems in which an electric current can flow. The thermal conductivity in the absence of electric current, ${\bf J}_E = 0$, which corresponds to a typical experimental situation, is given by
\begin{align}
\kappa
\equiv L_{TT} - \frac{L_{TE} L_{ET}}{L_{EE}}
= \frac{\mathcal{N}}{mT} \bigg( \langle\!\langle \xi_{\bf p}^2 \tilde{\tau}_{\bf p} \rangle\!\rangle - \frac{\langle\!\langle \xi_{\bf p} \tilde{\tau}_{\bf p} \rangle\!\rangle^2}{\langle\!\langle \tilde{\tau}_{\bf p} \rangle\!\rangle} \bigg),
\label{ThermalConductivity}
\end{align}
which is free from divergences in the clean limit.

Further discussions and illustrations of the main results obtained in this section for $\sigma$ of Eq.~\eqref{Sigma_EE}, $S$ of Eq.~\eqref{ThermoelectricPower}, and $\kappa$ of Eq.~\eqref{ThermalConductivity},
are presented in Sec.~\ref{Sec:Discussion} below.

\section{Beyond the relaxation-time approximation: Fermi liquid-type collision integral} 
\label{Sec:e-e collision integral beyond RTA}

In the previous section, we focused on the RTA for both collision integrals. We now improve our approach by modifying the electron-electron collision integral. Here, we will focus on the case of a constant scattering time $\tau_{ei}$, which allows us to transform the Boltzmann equation into a solvable form.

\subsection{Electron-electron collision integral}
\label{subsec:ee_coll_int}

We start by writing a refined form of the electron-electron collision integral as
\begin{align}
I_{ee}\{f\} 
& = - \frac{f_{\bf p}}{\tau_{out,{\bf p}}} + \frac{1 - f_{\bf p}}{\tau_{in,{\bf p}}},
\label{Coll_Int_ee_1}
\end{align}
where the out- and in-scattering rates are
\begin{align}
\frac{1}{\tau_{out,{\bf p}}}
& = \int_{{\bf q},{\bf p'},{\bf q'}} W_{{\bf p}{\bf q},{\bf p}'{\bf q}'} f_{\bf q} (1 - f_{{\bf p}'}) (1 - f_{{\bf q}'}),
\label{OutscatteringRate}
\\
\frac{1}{\tau_{in,{\bf p}}}
& = \int_{{\bf q},{\bf p'},{\bf q'}} W_{{\bf p}'{\bf q}',{\bf p}{\bf q}} f_{{\bf p}'} f_{{\bf q}'} (1 - f_{{\bf q}}).
\end{align}
The probability for electrons with momenta ${\bf p}, {\bf q}$ to be scattered into states with momenta ${\bf p}', {\bf q}'$ is \cite{remark_FL}
\begin{align}
W_{{\bf p}{\bf q},{\bf p}'{\bf q}'} 
& = s (2\pi)^{d+1} |U_{{\bf p}{\bf q},{\bf p}'{\bf q}'}|^2 \delta(\epsilon_{\bf p} + \epsilon_{\bf q} - \epsilon_{{\bf p}'} - \epsilon_{{\bf q}'}) 
\nonumber\\
& ~~~ \times \delta({\bf p} + {\bf q} - {\bf p}' - {\bf q}'),
\label{eq:W}
\end{align}
where $U_{{\bf p}{\bf q},{\bf p}'{\bf q}'}$ is the interaction matrix element.

Next, we write the deviation of the distribution function from the equilibrium one as 
\begin{align}
\delta f_{\bf p}
= - \Phi_{\bf p} \frac{\partial n_F(\xi_{\bf p})}{\partial \xi_{\bf p}} 
= \beta n_F(\xi_{\bf p}) [1 - n_F(\xi_{\bf p})] \Phi_{\bf p},
\label{DistributionFluctuation}
\end{align}
with $- \Phi_{\bf p}$ being the excess energy generated by the external perturbation. 
Inspired by Eqs.~\eqref{DistFluc_E}-\eqref{DistFluc_T}, we introduce the ansatz $
\Phi_{\bf p} 
= \sum_{\alpha\in\{E,T\}} \phi^\alpha(\xi_{\bf p}) \boldsymbol{v}_{\bf p} \cdot  {\bf F}_\alpha$ with ${\bf F}_E = - e {\bf E}$ and ${\bf F}_T = -\nabla_{\bf r} T$. The goal of our calculation is to find the two unknown functions $\phi^E(\xi_{\bf p})$ and $\phi^T(\xi_{\bf p})$. 
Finally, linearizing Eq.~\eqref{Coll_Int_ee_1} in $\Phi$, and applying the detailed balance principle to the equilibrium state, we find the canonical form of the linearized collision integral
\begin{align}
I_{ee}\{\Phi\}
& = - \frac{2(2\pi)^{d+1}}{T} \int_{{\bf q},{\bf p}',{\bf q}'} |U_{{\bf p}{\bf q},{\bf p}'{\bf q}'}|^2 
\nonumber\\
& ~~~ \times \delta(\epsilon_{\bf p} + \epsilon_{\bf q} - \epsilon_{{\bf p}'} - \epsilon_{{\bf q}'}) \delta({\bf p} + {\bf q} - {\bf p}' - {\bf q}') 
\nonumber\\
& ~~~ \times n_F(\xi_{\bf p}) n_F(\xi_{\bf q}) [1 - n_F(\xi_{{\bf p}'})] [1 - n_F(\xi_{{\bf q}'})] 
\nonumber\\
& ~~~ \times (\Phi_{{\bf p}} + \Phi_{{\bf q}} - \Phi_{{\bf p}'} - \Phi_{{\bf q}'}).
\label{Coll_Int_ee_2}
\end{align}

The collision integral given in Eq.~\eqref{Coll_Int_ee_2} is applied for any type of two-body interaction. In this work, we focus on two cases: (i) the FL interaction in three dimensions and (ii) the Coulomb interaction in two dimensions. For these cases, Eq.~\eqref{Coll_Int_ee_2} can be reduced to the same approximate form in the degenerate regime ($T \ll \epsilon_F$), as will be further discussed below. In order to formulate the result of this step, it is useful to separate $\phi^\alpha$ into symmetric and antisymmetric parts, $\phi^{\alpha}(\xi)=\phi^{\alpha}_s(\xi)+\phi^{\alpha}_a(\xi)$, where $\phi^{\alpha}_s(\xi)=\phi^{\alpha}_s(-\xi)$ and $\phi^{\alpha}_a(\xi)=-\phi^{\alpha}_a(-\xi)$. Then, the approximate form of the electron-electron collision integral reads
\begin{align}
& I_{ee}\{\phi\}
= - \frac{4n_F(\xi_{\bf p})[1-n_F(\xi_{\bf p})]}{\pi^2T^3\tau_{out}} \int_{-\infty}^\infty d\omega K(\omega,\xi_{\bf p})
\nonumber\\
& ~~~ \times \sum_{\alpha\in \{E,T\}} v_F \hat{\bf n}_{\bf p} \cdot {\bf F}_{\alpha} \sum_{\gamma\in\{s,a\}}[\phi^{\alpha}_\gamma(\xi_{\bf p}) - \Lambda_{\gamma}\phi^{\alpha}_\gamma(\xi_{\bf p}+\omega)],
\label{Coll_Int_ee_3}
\end{align} 
where we define $\hat{\bf n}_{\bf p} = {\bf p}/|{\bf p}|$, and  
\begin{align}
K(\omega,\xi_{\bf p}) 
= \omega n_B(\omega)\frac{1-n_F(\xi_{\bf p}+\omega)}{1-n_F(\xi_{\bf p})}.
\end{align}
Here, $n_{B}(\omega) = [\exp(\beta\omega) - 1]^{-1}$ is the Bose-Einstein distribution, $K(-\omega,-\xi_{\bf p}) = K(\omega,\xi_{\bf p})$, and $\int d\omega K(\omega, \xi_{\bf p}) = [\xi_{\bf p}^2+(\pi T)^2]/2$. In Eq.~\eqref{Coll_Int_ee_3}, $1/\tau_{out}$ denotes the out-scattering rate defined in  Eq.~\eqref{OutscatteringRate}, evaluated on the Fermi surface and in equilibrium. The dimensionless parameters $\Lambda_{s/a}$, relevant to the symmetric and antisymmetric parts of $\phi^\alpha$, respectively, depend on the interaction potential as well as on the dimensionality of the system. Both quantities, $1/\tau_{out}$ and $\Lambda_{s,a}$, will be specified below for the two model systems under consideration. Details of the derivation are described in Appendix~\ref{Appendix_Approximate_e-e_Coll}.

For the FL-type collision integral in three dimensions, one finds
\begin{align}
\Lambda_{s} = 1,
\qquad 
\Lambda_{a} = \frac{\displaystyle \bigg\langle \frac{|\tilde{U}(\theta,\varphi)|^2(1+2\cos\theta)}{\cos(\theta/2)}\bigg\rangle_{\rm av}}{\displaystyle \bigg\langle \frac{|\tilde{U}(\theta,\varphi)|^2}{\cos(\theta/2)}\bigg\rangle_{\rm av}},
\label{Lambda_a}
\end{align} 
Here, $\tilde{U}$ is obtained from $U$ by fixing all incoming and outgoing momenta to $p_F$. Two angles are used for characterizing $\tilde{U}$: $\theta$ is the angle between the two incoming momenta and $\varphi$ is the angle between the two planes spanned by the incoming momenta and by the outgoing momenta. The angular average is defined as $\langle X(\theta,\varphi) \rangle_{\rm av} = \int_0^{2\pi} d(\varphi/2\pi) \int_0^\pi\sin\theta d\theta X(\theta,\varphi)$. It follows that $\Lambda_a$ can take any value between $-1$ and $3$. The lowest value, $\Lambda_a=-1$, corresponds to head on collisions, $\tilde{U}(\theta=\pi,\varphi)$, and the highest value, $\Lambda_a=3$ to collinear scattering, $\tilde{U}(\theta=0,\varphi)$.

The out-scattering rate increases with temperature as $T^2$, as is characteristic for FLs
\begin{align}
\frac{1}{\tau_{out}}
= u \frac{T^2}{\epsilon_F},
\label{tau_out_3D}
\end{align}
where we define the dimensionless parameter
\begin{align}
u = \frac{m^3\epsilon_F}{32\pi} \bigg\langle\frac{|\tilde{U}(\theta,\varphi)|^2}{\cos(\theta/2)}\bigg\rangle_{\rm av}.
\label{eq:u}
\end{align}

For the Coulomb interaction in two dimensions, we find the simple result
\begin{align}
\Lambda_s = \Lambda_a = 1.\label{eq:LambdaC}
\end{align}
Our result for $\Lambda_a$ agrees with the value found in Ref.~\cite{Lyakhov03}, where the thermal conductivity of a clean two-dimensional electron gas was studied. The value of $\Lambda_a$ for the Coulomb interaction in two dimensions falls within the range available for the three-dimensional FL. The only difference between the two collision integrals is in the out-scattering rate, which for the two-dimensional case takes the form
\begin{align}
\frac{1}{\tau_{out}}
= \frac{\pi T^2}{8 \epsilon_F} \ln \bigg|\frac{4\epsilon_F}{T}\bigg|.
\label{tau_out_2D}
\end{align}
In the derivation of the collision integral and the scattering rate, we assumed that the main contribution to the collision integral arises from forward scattering. This can be justified for small-$r_s$ systems, for which the random phase approximation is applicable~\cite{Jungwirth96,Lyakhov03}. The low dimensionality limits the available phase space for collisions \cite{Hodges71}. As a result, the out-scattering rate differs from the three-dimensional case by a logarithmic correction.

We emphasize that the renormalization of  the collision amplitude to disorder has not been accounted for in the derivation. This limits the range of applicability of Eqs.~\eqref{Coll_Int_ee_3}, \eqref{eq:LambdaC}, and \eqref{tau_out_2D}

\subsection{Non-equilibrium distribution and conductivities}
\label{Sec:Non-equilibrium distribution}

The solution of the linearized Boltzmann equation with electron-electron collision integral given by Eq.~\eqref{Coll_Int_ee_3}, and electron-impurity collision integral in the RTA, $I_{ei}=-\delta f/\tau_{ei}$ with $\delta f$ given by Eq.~\eqref{DistributionFluctuation}, was found using the method introduced in Refs.~\cite{Brooker68} and \cite{Jensen68} for finding transport coefficients of a clean three-dimensional FL, and generalized to include disorder in Ref.~\cite{Bennett69}. The method essentially amounts to diagonalizing the collision integral. This is achieved by mapping the Boltzmann equation via Fourier transform to an inhomogeneous second-order differential equation. In the differential equation, the inhomogeneity arises from the driving term of the Boltzmann equation. The associated homogeneous equation resembles the Schr\"odinger equation for a particle in a $\mbox{sech}^2x$ potential, and can be solved with the help of an eigenfunction expansion. This expansion is also the key to solving the inhomogeneous equation and to the calculation of transport coefficients. For the convenience of the reader, we summarize the main steps in Appendix \ref{Appendix:solvingFLKE}.

We further simplified the solution for the thermal conductivity found in Ref.~\cite{Bennett69}, bringing it to a form that is suitable for numerical evaluation
\begin{align}
\frac{\kappa^{(0)}}{\kappa_0} 
& = \frac{\tau_{out}}{\tau_{ei}} \sum_{n=0}^\infty 
\frac{3(2n + \varepsilon + 3/2)}{8[\lambda_{2n+1}(\varepsilon) - \Lambda_a]} \frac{\Gamma(n+3/2)\Gamma(n+\varepsilon+3/2)}{\Gamma(n+1)\Gamma(n + \varepsilon + 1)}
\nonumber\\
& ~~~ \times \frac{[\Gamma(n+(\varepsilon+1)/2)]^2}{[\Gamma(n+\varepsilon/2+2)]^2}.
\label{Sigma_TT_Refined}
\end{align}
Here, $\kappa_0 = \pi^2 \mathcal{N} T \tau_{ei} / (3m)$, $\lambda_n(\varepsilon) = {(n + \varepsilon) (n + \varepsilon + 1)}/{2}$ and $\varepsilon=\sqrt{1+\tau_{out}/(2\tau_{ei})}$. The thermoelectric transport coefficients are smaller than the electric and thermal conductivities by a factor of $T/\epsilon_F$. Therefore, in the FL approximation, there is no difference between the thermal conductivity in an open system where current can flow, $L_{TT}$, and in a closed system with ${\bf J}_{E}=0$, $\kappa=L_{TT}-L_{TE}L_{ET}/L_{EE}$.

One can derive an expression similar to Eq.~\eqref{Sigma_TT_Refined} for the electric conductivity; see Appendix \ref{Appendix:solvingFLKE} and \ref{Appendix_InverseFourierTransform}. However, as already mentioned in Sec.~\ref{Subsec:NED}, we know that $\sigma=\sigma_0$ is unaffected by electron-electron collisions for the case of constant electron-impurity scattering, and therefore $\phi_s^E(\xi_{\bf p}) = \tau_{ei}$. This is also consistent with Eq.~\eqref{DistFluc_E_ConstScatt} in the RTA. This simple result can be used 
for a consistency check of the eigenfunction decomposition obtained from mapping to the Schr\"odinger equation; see Appendix \ref{Appendix:ElectricConductivity}. In Appendices \ref{Appendix:solvingFLKE} and \ref{Appendix_InverseFourierTransform}, we also derive the following expression characterizing the deviation of the distribution function from the equilibrium distribution due to a temperature gradient,
\begin{align}
\phi_a^T(\xi_{\bf p})
= \frac{\pi^2}{2} \tau_{out} {\rm cosh}\bigg(\frac{\beta\xi_{\bf p}}{2}\bigg) 
\sum_{n=0}^\infty 
C_{2n+1}^T Q_{2n+1}(\xi_{\bf p}).
\label{NoneqDistFluc_Refined1}
\end{align}
Here, we defined
\begin{align}
& Q_n(\xi_{\bf p}) 
= \frac{2^{\varepsilon-2}}{\pi^2} \frac{|\Gamma(\varepsilon/2 + i\beta\xi_{\bf p}/(2\pi))|^2}{\Gamma(\varepsilon)}
\nonumber\\
& ~~~~~~ \times\!_3F_2\left(
\begin{array}{c}
-n,n+2\varepsilon+1,\varepsilon/2+i\beta\xi_{\bf p}/(2\pi)
\\
\varepsilon+1,\varepsilon
\end{array}
; 1\right),
\label{InverseFourierTransformQn}
\end{align}
with $_3F_2$ being the generalized hypergeometric function. A more detailed analysis of Eq.~\eqref{NoneqDistFluc_Refined1} is presented in the following section.

\section{Discussion}
\label{Sec:Discussion}

In the two previous sections, we derived formulas for the non-equilibrium distribution functions and conductivities in two different approaches. In this section, we present a more detailed analysis of both approaches, compare, and illustrate them. In Sec.~\ref{Sec:Discussion:Momentum-independent}, we discuss properties of conductivities for constant electron-impurity scattering time, $\tau_{ei,{\bf p}}=\tau_{ei}$. In this case, we can compare the results obtained by the two approaches, the RTA and the eigenfunction expansion, for the electric and thermal conductivities as well as the corresponding distribution functions. Then, in Sec.~\ref{Sec:Discussion:Momentum-dependent}, we analyze the influence of momentum-dependent electron-impurity scattering on the transport coefficients on the basis of the RTA.

The model of two independent terms in the collision integral used in this paper allows us to discuss in detail how elastic and inelastic collisions interfere with each other when calculating transport coefficients. However, it is too simplistic to cover all aspects of the complicated interplay of disorder and interactions, even when renormalizations of parameters are accounted for. A well-known example is the violation of the  Wiedemann-Franz law in the disordered electron liquid that has been extensively studied in the diffusive \cite{Raimondi04,Catelani05,Niven05,Schwiete16a,Schwiete16b} and ballistic transport regimes \cite{Catelani05} (under the assumption that $\tau_{ee}\gg\tau_{ei}$). The inclusion of this effect would require a fully microscopic treatment that is beyond the scope of this paper. With this reservation in mind, we now turn to the results obtained in the framework of our model.

\subsection{Momentum-independent electron-impurity scattering}
\label{Sec:Discussion:Momentum-independent}

The electric/thermal conductivities [Eqs.~\eqref{Sigma_EE}, \eqref{ThermalConductivity}] and the Seebeck coefficient [Eq.~\eqref{ThermoelectricPower}] derived using the method of Keyes acquire a simple form when the scattering rate by impurities is constant~\cite{Principi15,Lucas18,Remark_kappa_hyd}
\begin{align}
\sigma^{(0)}
& = \frac{\mathcal{N}e^2\tau_{ei}}{m},
\label{ElectricConductivity_0}
\\
\kappa^{(0)}
& = \frac{\mathcal{N}}{mT} \frac{\langle\!\langle \xi_{\bf p}^2 \rangle\!\rangle - \langle\!\langle \xi_{\bf p} \rangle\!\rangle^2}{\tau_{ei}^{-1} + \tau_{ee}^{-1}},
\label{ThermalConductivity_0}
\\
S^{(0)}
& = - \frac{\langle \!\langle \xi_{\bf p}\rangle\!\rangle}{eT}.
\label{Thermoelectricpower_0}
\end{align}
Out of these three coefficients, only $\kappa$ depends on $\tau_{ee}$ in the absence of a momentum dependence of $\tau_{ei}$.
It is governed by the Matthiessen's rule ($\sim \tau_{ei}^{-1} + \tau_{ee}^{-1}$) for two scattering rates \cite{Principi15,Lucas18}. For FLs at low temperatures $T\ll \epsilon_F$, the moments of $\xi_{\bf p}$ that enter the transport coefficients become $\langle\!\langle \xi_{\bf p}^2 \rangle\!\rangle = \pi^2 T^2/3$ and $\langle\!\langle \xi_{\bf p} \rangle\!\rangle = \pi^2T^2/(3\epsilon_F)$.

The only difference between the conductivities in Eqs.~\eqref{Sigma_EE}-\eqref{ThermalConductivity} and those obtained using the eigenfunction expansion is the electron-electron collision integral. In the RTA, $\tau_{ee}$ is independent of momentum. In the FL case, the out-scattering rate at finite $\xi_{\bf p}$ is $\tau^{-1}_{out,{\bf p}}\propto \int d\omega K(\omega,\xi_{\bf p}) = (\xi_{\bf p}^2 + \pi^2 T^2)/2$. In a somewhat simplified form, the question addressed in this section may therefore be stated as follows: How does the $\xi_{\bf p}$ dependence of the out-scattering rate influence the transport coefficients?

\subsubsection{Characteristics of the electric conductivity}

As argued before, for the system we study, the non-equilibrium distribution in the presence of an applied electric field is independent of the electron-electron collisions when the electron-impurity scattering is momentum-independent \cite{Remark_Fermi_surface}. As a result, $\sigma^{(0)}$ is equal to the Drude conductivity $\sigma_0 = \mathcal{N}e^2\tau_{ei}/m$ at all temperatures, and $\phi_s^E(\xi_{\bf p})=\tau_{ei}$. The momentum dependence of the out-scattering rate is insignificant in this case. This result was confirmed explicitly for the RTA in Sec.~\ref{Sec:Keyes approach} and also for the FL-type collision integral in Sec.~\ref{Sec:e-e collision integral beyond RTA} and Appendix \ref{Appendix:ElectricConductivity}.

\subsubsection{Characteristics of the thermal conductivity}

As we mentioned in Sec.~\ref{Sec:Keyes approach} B, the underlying mechanism for thermal driving is fundamentally different from electric driving, because the driving force in the former case is proportional to $\xi_{\bf p}$, loading different weights on different energy shells; see Eq.~\eqref{BoltzmannEq_Linearized}. As a result, the non-equilibrium distribution, and thus the thermal conductivity, depends on the electron-electron collisions even for constant electron-impurity scattering time, and violates the Wiedemann-Franz law. The Matthiessen's rule works in the RTA, while it is mildly violated for the more complex collision integral as will be shown below.

In Fig.~\ref{Figure1}(a), we plot the thermal conductivity $\kappa^{(0)}$ of a two-dimensional electron system found using the RTA, Eq.~\eqref{ThermalConductivity_0}, and from the eigenfunction expansion for the Coulomb interaction. 
In this figure, the analog of $\sigma_0$ for the thermal conductivity, $\kappa_0=\pi^2 \mathcal{N}T\tau_{ei}/(3m)$, is used for the normalization of $\kappa^{(0)}$. Any deviation of $\kappa^{(0)}$ from $\kappa_0$ is tantamount to a violation of the Wiedemann-Franz law. Indeed, the relations $\sigma=\sigma_0$ and $\kappa=\kappa^{(0)}$ hold in the present situation. It follows that $\kappa^{(0)} / \kappa_0 = \mathcal{L} / \mathcal{L}_0$, where $\mathcal{L}_0 \equiv \kappa_0/(\sigma_0T) = \pi^2/(3e^2)$ is the Lorenz number, and $\mathcal{L} = \kappa / (\sigma T)$ is the generalized Lorentz number.

To plot Eq.~\eqref{ThermalConductivity_0}, we need to know the inelastic scattering rate. This is because in the RTA, $\tau_{ee}$ simply plays the role of a phenomenological parameter, while $\tau_{out}$ in the eigenfunction expansion is derived from the microscopic theory. Here, and in the rest of the discussion, we fix $\tau_{ee}$ by matching the thermal conductivities in the clean limit ($\tau_{ei} \gg \tau_{ee}, \tau_{out}$). A detailed explanation of the procedure is provided in Appendix~\ref{Appendix:BC}. In the following discussion, based on Eqs.~\eqref{Relation_tau_ee_2D} and \eqref{Relation_tau_ee_3D}, we use $\tau_{out}$ instead of $\tau_{ee}$.

\begin{figure}[t]
\centering
\includegraphics[width=0.46\textwidth]
{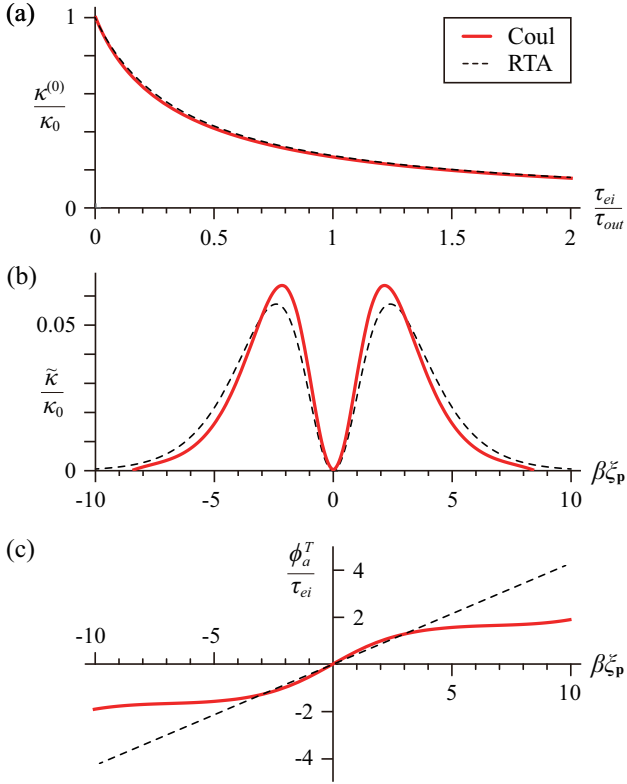} \\
\caption{
(a) Thermal conductivity $\kappa^{(0)}$ measured in units of $\kappa_0 = \pi^2 \mathcal{N} T \tau_{ei} / (3m)$ as a function of $\tau_{ei}/\tau_{out}$. The curves were produced by solving the Boltzmann equation with the eigenfunction-expansion approach for Coulomb interactions in two dimensions (solid line), and using Keyes method within the RTA (dashed line). The inelastic relaxation time $\tau_{ee}$ was chosen such that both results match in the clean limit (see discussion in Sec.~\ref{Sec:Discussion:Momentum-independent}). In panel (b) and (c), we plot the corresponding $\tilde{\kappa}(\beta\xi_{\bf p})/\kappa_0$ and $\phi_a^T(\beta\xi_{\bf p})/\tau_{ei}$. $\tilde{\kappa}(\beta\xi_{\bf p})$ encodes the contribution from different energy shells. Here, we set $\tau_{ei}/\tau_{out} = 0.5$.
}
\label{Figure1}
\end{figure}

\begin{figure}[t]
\centering
\includegraphics[width=0.46\textwidth]
{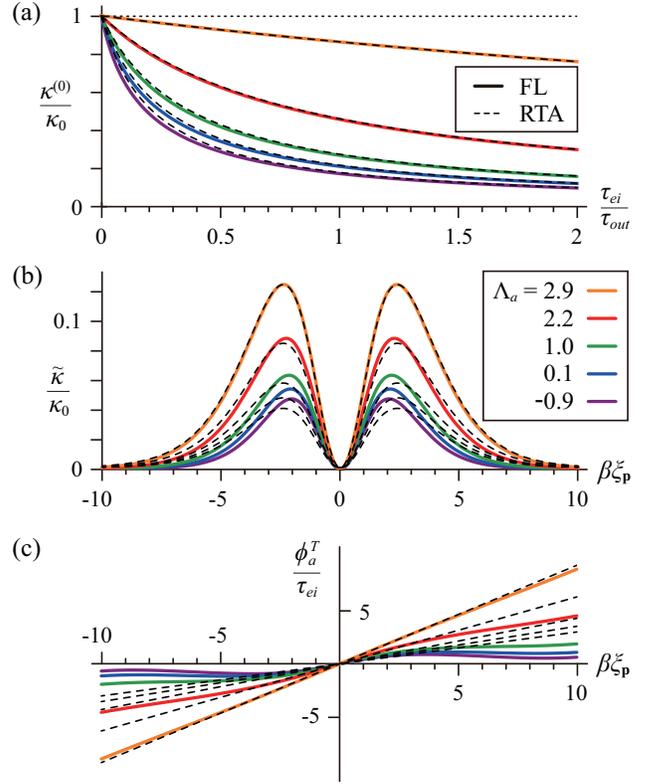} \\
\caption{
The counterpart of Fig.~\ref{Figure1} for FL interactions in three dimensions. Each solid line represents a calculation for inelastic scattering with different angular dependence of the scattering probability, i.e., for various values of $\Lambda_a$. The dashed lines show the corresponding results for the thermal conductivity calculated using the RTA. For the RTA, we used the relation $1/\tau_{ee}=C(\Lambda_a)T^2/\varepsilon_F$ to fix the inelastic relaxation time; $C(\Lambda_a)$ is a temperature-independent constant defined in Appendix \ref{Appendix:BC}. The dotted black line is the solution for $\Lambda_a\rightarrow3$, where inelastic collisions do not relax the thermal current. In panel (b) and (c), we set $\tau_{ei}/\tau_{out} = 0.5$.
}
\label{Figure2}
\end{figure}

We observe that $\kappa^{(0)}/\kappa_0$ obtained from the two different models are fairly close to each other. Both predict a strong suppression of the thermal conductivity due to frequent electron-electron collisions at elevated temperatures. Only a slight mismatch between two results is observed for $0 \lesssim \tau_{ei}/\tau_{out} \lesssim 1$.

The excellent agreement between the results for the thermal conductivity obtained by the two different models is very surprising. It stands in contrast to the observation that electron-electron collisions play an important role in relaxing thermal currents. In particular, since the force generated by a temperature gradient is non-uniform in momentum, we expected the thermal conductivity to be sensitive to the exact momentum dependence of the inelastic scattering, and then the two different electron-electron collision integrals would result in significantly different $\kappa^{(0)}$. In both cases, the thermal conductivity is determined by integrating
the non-equilibrium distribution function over different
energy shells, $\kappa^{(0)}=\int_{-\infty}^{\infty} dx \tilde{\kappa}_i(x)$, where $x=\beta \xi_{\bf p}$, and the label $i\in\{\rm RTA, Coul\}$ distinguishes the RTA and the Coulomb collision integral. For the RTA, under the approximation of momentum-independent electron-impurity scattering, the integral is incorporated in the averages $\left\langle\!\left\langle \dots \right\rangle\!\right\rangle$ in Eq.~\eqref{ThermalConductivity_0}. At low temperatures, where the first term in the numerator is dominant, this equation takes the approximate form 
\begin{align}
\kappa^{(0)}
= - \frac{2\nu_2\mu T}{m(\tau_{ei}^{-1}+\tau_{ee}^{-1})}\int dx  x^2 \partial_x n_F(x),
\end{align}
where $\nu_2= m/(2\pi)$ is the density of states at the Fermi energy in two dimensions. The thermal conductivity calculated using the eigenfunction expansion can be written in a similar form; see Eq.~\eqref{Sigma_x}. In Fig.~\ref{Figure1}(b), we compare $\tilde{\kappa}_{\rm RTA}$ and $\tilde{\kappa}_{\rm Coul}$ for $\tau_{ei}/\tau_{out}=0.5$. A key observation is that the structure of the two integrands is comparable; they both have a peak between $x=2$ and $3$ and decay exponentially for $x\gtrsim 5$. The smallness at large $x$ stems from the derivative of the Fermi-Dirac distribution that enters $\tilde{\kappa}$ in both cases. Likewise, the two integrands vanish as $x\rightarrow 0$ due to the energy $\xi_{\bf p}$ that appears in the definition of the thermal current and another factor $\xi_{\bf p}$ that originates from the thermal force.

\begin{figure}[t]
\centering
\includegraphics[width=0.46\textwidth]
{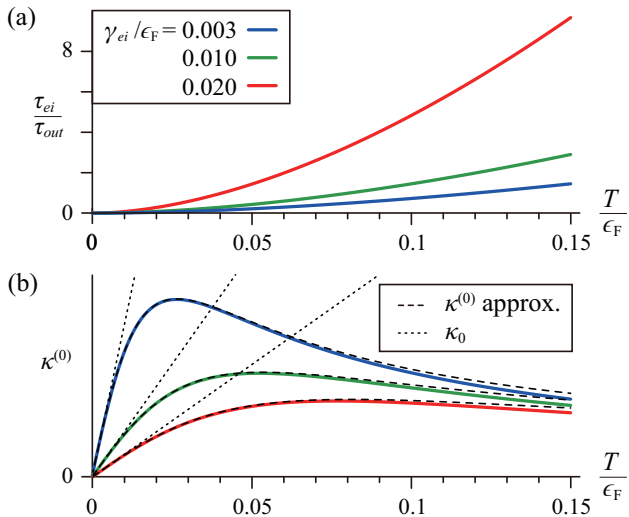} \\
\caption{Temperature dependence of $\tau_{ei}/\tau_{out}$ at the Fermi energy $\epsilon_F$ [panel (a)], and  the thermal conductivity $\kappa^{(0)}$ [panel (b)]. The latter is shown in arbitrary units.  All curves were obtained treating the electron-electron collision integral within the RTA in two dimensions. Each line corresponds to different values of $\gamma_{ei} = \tau_{ei}^{-1}$. In panel (b), the solid lines represent the full solution, while the dashed black lines were obtained through a perturbative expansion in $T/\epsilon_F$. The dotted black lines show $\kappa_0$ for the same values of $\tau_{ei}$.}
\label{Figure3}
\end{figure}

\begin{figure}[t]
\centering
\includegraphics[width=0.46\textwidth]
{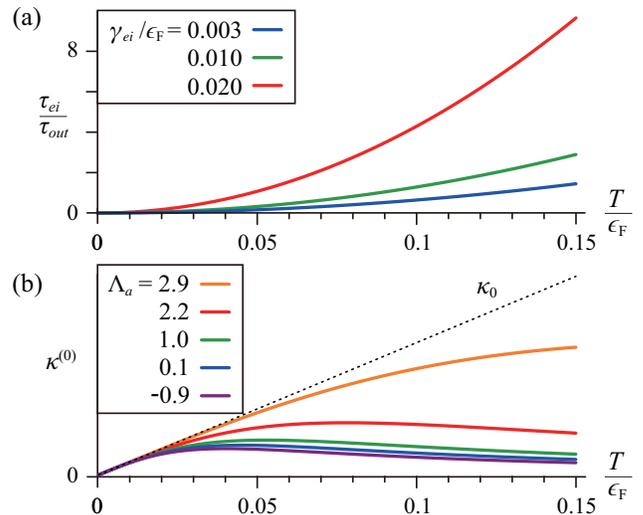} \\
\caption{The  counterpart to Fig.~\ref{Figure3} for FL interactions in three dimensions, with $u=1.29$. In panel (b), we fix $\gamma_{ei}/\epsilon_F = 0.01$ and look at different values of $\Lambda_a$.}
\label{Figure4}
\end{figure}

To better understand the similarities and differences between the two methods, we recall that the integrands are given by $\tilde{\kappa}_{i}\propto \phi_a^{T,i}(x) x\partial_xn_F(x)$ (see Appendix~\ref{Appendix:solvingFLKE}). This allows us to factor out common terms and to focus on $\phi_a^{T,i}(\xi_{\bf p})$, which is defined in Eq.~\eqref{DistributionFluctuation}, and incorporates the deviation of the distribution function from the one at equilibrium. In Fig.~\ref{Figure1}(c), we plot $\phi_a^T$ that is calculated first using the RTA and second with Coulomb collision integral. We obtain that for $x\lesssim2$ the two distribution functions are approximately linear in $x$ with almost identical slope. The slope is fixed by the matching procedure for the inelastic scattering times $\tau_{ee}$ and $\tau_{out}$ [recall that $\phi_a^{T,{\rm RTA}}=x/(\tau_{ei}^{-1}+\tau_{ee}^{-1})$]. The two curves for $\phi_a^T$ in Fig.~\ref{Figure1}(c) strongly deviate at larger values of $x$. However, the main contribution to the integral over $x$ arises from small $x$ due to the derivative of the Fermi-Dirac distribution. Consequently, the thermal conductivity is only weakly sensitive to the nature of the inelastic collisions. Moreover, we recall that for the Coulomb interaction $\tau^{-1}_{out,{\bf p}}\propto \xi_{\bf p}^2 + \pi^2T^2$ (up to slowly varying logarithmic factors). Hence, the out-scattering time is roughly constant for $x\lesssim\pi$, and the Boltzmann equations used in the two methods for finding the non-equilibrium distribution are almost identical in this range of energies, once $\tau_{ee}$ is appropriately fixed.

Since $\tau^{-1}_{out,{\bf p}}\propto \xi_{\bf p}^2 + \pi^2T^2$ also holds for FL interactions in three dimensions, we expect that the RTA gives a very good approximation for the thermal conductivity in this case as well. In Fig.~\ref{Figure2}, we compare the thermal conductivity found using the RTA and the eigenfunction expansion for FL interactions in three dimensions. For this purpose, we repeat the steps performed in Fig.~\ref{Figure1}. In the FL case, the parameter $\Lambda_{a}$ changes between $-1$ and $3$  depending on the angular dependence of the collisions; in Fig.~\ref{Figure2}, we display curves for different values of $\Lambda_{a}$. We observe that for $\Lambda_a=3$, the thermal conductivity $\kappa^{0}$ is constant. This is a manifestation of the fact that collinear scattering cannot relax the thermal current. As the angle between colliding electrons grows, i.e., $\Lambda_{a}$ decreases, the thermal conductivity decays with increasing $1/\tau_{out}$. At the same time, the difference between the thermal conductivities obtained by the RTA and from the Coulomb collision integral grows, despite being overall small. Finally, we note that the results for the three-dimensional FL with $\Lambda_a=1$ and for the two-dimensional Coulomb-interaction coincide, as was discussed in Sec.~\ref{subsec:ee_coll_int}.

The temperature dependence of the thermal conductivities is implicit in Figs.~\ref{Figure1}(a) and \ref{Figure2}(a). This is because plotting $\kappa^{(0)}/\kappa_0$ removes the explicit temperature dependence through $\kappa_0$, and plotting the thermal conductivities as a function of $\tau_{ei}/\tau_{out}$ accounts for the temperature dependence of the out-scattering rate $1/\tau_{out}$.  The only restriction on temperature is related to the validity of the FL approximation that assumes $T\ll\epsilon_F$. By contrast, the RTA can be valid at any temperature. In practice, we use the result obtained from the eigenfunction expansion to fix the inelastic scattering rate for the RTA. To complete our analysis, in Fig.~\ref{Figure3}, we use the temperature dependence of $1/\tau_{out}$ for the Coulomb interaction [Eq.~\eqref{tau_out_2D}] to plot $\kappa^{(0)}(T)$ found using the RTA. Similarly, in Fig.~\ref{Figure4}, we show $\kappa^{(0)}(T)$ using the inelastic relaxation rate obtained for the three-dimensional FL [Eq.~\eqref{tau_out_3D}].

\subsubsection{Characteristics of the Seebeck coefficient}

Just like the electric conductivity, it is understood that the Seebeck coefficient $S^{(0)}$ is not modified by electron-electron collisions within the RTA. For FLs, in which $T\ll\epsilon_F$, we estimate $S^{(0)} \propto T/\epsilon_F$ from Eq.~\eqref{Thermoelectricpower_0}. Note, however, that thermoelectric power is most sensitive to system details among the transport coefficients. The reason is that $S^{(0)}$ is small but non-zero only in the presence of particle-hole asymmetry, an effect beyond the framework of FL theory. The particle-hole asymmetry can be caused, for example, from a non-constant electron velocity or density of states, or from the dependence of the elastic scattering rate on momentum. In the estimate presented for $S^{(0)}$ above, a finite result was obtained 
due to a non-constant velocity in both dimensions as well as the non-constant density of state in three dimensions, which yields $\langle\!\langle \xi_{\bf p}\rangle\!\rangle \neq 0$. Section \ref{Sec:Seebeck} is devoted to a more detailed discussion of $S$.

\subsection{Momentum-dependent electron-impurity scattering}
\label{Sec:Discussion:Momentum-dependent}

We turn to analyzing the conductivities for a general momentum-dependent elastic scattering term, based on the RTA. We start by studying the electric conductivity in the two limits when either elastic or inelastic scattering dominates. As in Ref.~\cite{Keyes58}, we find 
\begin{align}
& \sigma
= \frac{\mathcal{N}e^2}{m}
\left\{
\begin{array}{ll}
\langle\!\langle \tau_{ei,{\bf p}}\rangle\!\rangle,
& \quad \tau_{ee}\gg \tau_{ei,{\bf p}} \vspace{0.2cm}
\\
\langle\!\langle \tau^{-1}_{ei,{\bf p}}\rangle\!\rangle^{-1},
& \quad \tau_{ee}\ll \tau_{ei,{\bf p}}.
\end{array}
\right.
\end{align}
Although $\tau_{ee}$ drops out in both limits, this result clearly demonstrates that the electric conductivity is modified by electron-electron collisions for a momentum-dependent electron-impurity scattering time.

Similarly, in the two limits, the thermal conductivity reads
\begin{align}
& \kappa
= \frac{\mathcal{N}}{mT}
\left\{
\begin{array}{cl}
\displaystyle
\langle\!\langle \xi^2_{\bf p}\tau_{ei,{\bf p}}\rangle\!\rangle - \frac{\langle\!\langle \xi_{\bf p}\tau_{ei,{\bf p}}\rangle\!\rangle^2}{\langle\!\langle \tau_{ei,{\bf p}}\rangle\!\rangle},
& ~~  \tau_{ee}\gg \tau_{ei,{\bf p}} \vspace{0.2cm}
\\
\tau_{ee}\left(\langle\!\langle \xi_{\bf p}^2\rangle\!\rangle-\langle\!\langle \xi_{\bf p}\rangle\!\rangle^2\right),
& ~~ \tau_{ee}\ll \tau_{ei,{\bf p}}.
\end{array}
\right.
\end{align}
In these limits, $\kappa$ is determined by the dominant scattering process, since both elastic and inelastic scattering can relax thermal currents. Unlike the electric conductivity, the thermal conductivity remains finite in the absence of impurities. The results for $\sigma$ and $\kappa$ in the limit of frequent electron-electron collisions suggest a strong violation of the Wiedemann-Franz law.  Obviously, the momentum dependence of $\tau_{ei,{\bf p}}$ has no influence on the result for $\kappa$ in this limit. This is different in the disorder-dominated limit, where the momentum dependence of $\tau_{ei,{\bf p}}$ can play an important role.

For the Seebeck coefficient, we find from Eq.~\eqref{ThermoelectricPower}
\begin{align}
S = - \frac{1}{eT}
\left\{
\begin{array}{cl}
\langle\!\langle \xi_{\bf p}\tau_{ei,{\bf p}}\rangle\!\rangle/\langle\!\langle \tau_{ei,{\bf p}}\rangle\!\rangle,
& \quad \tau_{ee}\gg \tau_{ei,{\bf p}}
\\
\langle\!\langle \xi_{\bf p}\rangle\!\rangle,
& \quad \tau_{ee}\ll \tau_{ei,{\bf p}}.
\end{array}
\right.
\end{align}
We see that, similar to the thermal conductivity, the momentum dependence of $\tau_{ei,{\bf p}}$ can be important in the disorder-dominated limit, while it has no influence in the electron-electron collision dominated regime.

Next, we study pertubatively the effect of a momentum-dependent elastic scattering rate on the transport coefficients. Therefore, we separate the electron-impurity scattering rate into two parts:
\begin{align}
\frac{1}{\tau_{ei,{\bf p}}}
= \frac{1}{\tau_{ei}} + \delta\Gamma_{\bf p}.
\label{e-impRate_separation}
\end{align}

We expand the electric/thermal conductivities [Eqs.~\eqref{Sigma_EE}, \eqref{ThermalConductivity}] and the Seebeck coefficient [Eq.~\eqref{ThermoelectricPower}] up to linear order in $\delta \Gamma_{\bf p}$. The zeroth-order terms have already been displayed in Eqs.~\eqref{ElectricConductivity_0}-\eqref{Thermoelectricpower_0}. The first-order corrections in $\delta \Gamma_{\bf p}$ take the form
\begin{align}
& \delta \sigma^{(1)}
= - \frac{\mathcal{N}e^2\tau_{ei}^2}{m} \langle\!\langle \delta\Gamma_{\bf p} \rangle\!\rangle,\label{ElectricConductivity_Keyes_1st_Order}
\\
& \delta \kappa^{(1)}
= \frac{\mathcal{N}}{mT}\frac{2 \langle\!\langle \xi_{\bf p} \rangle\!\rangle \langle\!\langle \xi_{\bf p} \delta\Gamma_{\bf p} \rangle\!\rangle-\langle\!\langle \xi_{\bf p} \rangle\!\rangle^2 \langle\!\langle \delta\Gamma_{\bf p} \rangle\!\rangle - \langle\!\langle \xi_{\bf p}^2 \delta\Gamma_{\bf p} \rangle\!\rangle}{(\tau_{ei}^{-1} + \tau_{ee}^{-1})^2},\label{ThermalConductivity_Keyes_1st_Order}
\\
& \delta S^{(1)}
= \frac{1}{eT}\frac{\langle\!\langle \xi_{\bf p}\delta \Gamma_{\bf p}\rangle\!\rangle - \langle\!\langle\xi_{\bf p}\rangle\!\rangle\langle\!\langle\delta\Gamma_{\bf p}\rangle\!\rangle}{\tau_{ei}^{-1}+\tau_{ee}^{-1}}.
\label{ThermoelectricPower_Keyes_1st_Order}
\end{align}
For the electric conductivity, both the zeroth- and first-order terms in $\delta \Gamma_{\bf p}$ are independent of $\tau_{ee}$. In fact, electron-electron collisions affect $\sigma$ starting from the second order only. As expected, $\tau_{ee}$ enters both $\kappa^{(0)}$ and $\delta \kappa^{(1)}$. Interestingly, while the leading term, $S^{(0)}$, has no dependence on scattering times, the correction $\delta S^{(1)}$ reflects the interplay of both scattering times, $\tau_{ei}$ and $\tau_{ee}$. The momentum dependence of $\tau_{ei}$ therefore induces a sensitivity of $S$ to electron-electron collisions already at the first order in $\delta \Gamma_{\bf p}$, unlike for the electric conductivity. Next, we examine the corrections to the conductivities and the Seebeck coefficient in the presence of two types of momentum-dependent elastic scattering rates.

\subsubsection{Analytic form of the elastic scattering rate}

We first consider an electron-impurity scattering rate that is an analytical function of energy. Then, we can expand $\delta \Gamma_{\bf p}$ as a power series in $\xi_{\bf p}$,
\begin{align}
\frac{\delta\Gamma_{\bf p}}{\gamma_{ei}} 
= \sum_{n\in\mathbb{N}} w_n \bigg(\frac{\xi_{\bf p}}{\epsilon_F}\bigg)^n,
\end{align}
where we define the constant electron-impurity scattering rate $\gamma_{ei} = \tau_{ei}^{-1}$, and the expansion coefficient $w_n$. The corrections to the transport coefficients are then easily found with the help of the general formulas derived in the previous section. For the purpose of illustration, we assume $w_n \ne 0$ only for $n=1,2$.

\begin{figure}[t]
\centering
\includegraphics[width=0.46\textwidth]
{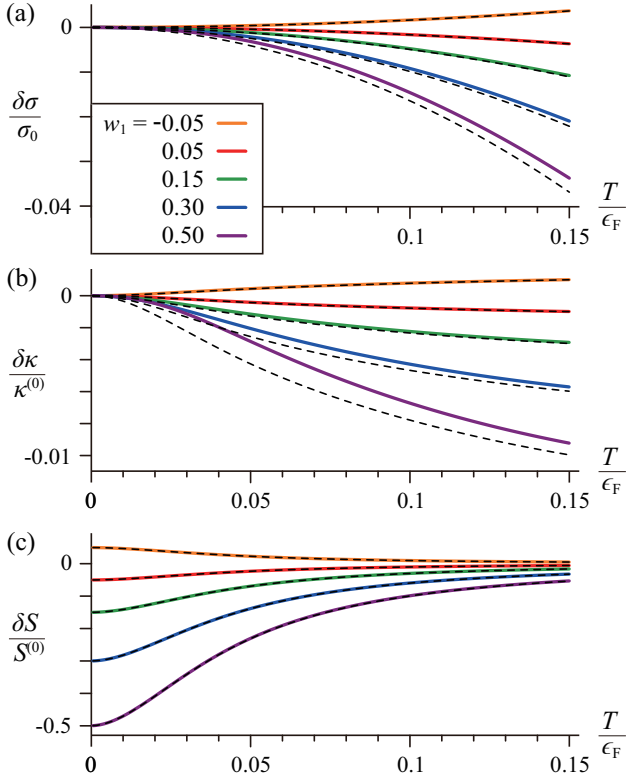} \\
\caption{
The relative deviations of the electric conductivity $\delta\sigma/\sigma^{(0)}$ [panel (a)], thermal conductivity $\delta\kappa/\kappa^{(0)}$ [panel (b)] and Seebeck coefficient $\delta S/S^{(0)}$ [panel (c)] caused by a momentum dependent elastic scattering rate, $1/\tau_{ei,\mathbf{p}}=\gamma_{ei}+\delta\Gamma_{\bf p}$ where  
$\delta\Gamma_{\bf p} = w_1 \gamma_{ei}\xi_{\bf p}/\epsilon_F$. All the curves were calculated within the RTA, with the relaxation rate matched to the one for two-dimensional Coulomb interactions (Sec.~\ref{Sec:Discussion:Momentum-independent}). Each curve corresponds to a different value of $w_1$, and $\gamma_{ei}/\epsilon_F = 0.01$. The dashed black lines show the results of the perturbative expansion with respect to  $w_1$. 
}
\label{Figure5}
\end{figure}

Then, the corrections to the conductivities at low temperatures ($T\ll \epsilon_F$) become (in dimensions $d=2,3$)
\begin{align}
\frac{\delta \sigma^{(1)}}{\sigma_0} 
& = - \bigg(\frac{1}{6} d w_1 + \frac{1}{3} w_2\bigg) \bigg(\frac{\pi T}{\epsilon_F}\bigg)^2,
\label{AnalyticCorrection_ElectricConductivity}
\\
\frac{\delta \kappa^{(1)}}{\kappa^{(0)}}
& = - \bigg(\frac{11}{30} d w_1 + \frac{7}{5} w_2 \bigg) \frac{(\pi T/\epsilon_F)^2}{1 + \tau_{ei}/\tau_{ee}},
\label{AnalyticCorrection_ThermalConductivity}
\\
\frac{\delta S^{(1)}}{S^{(0)}}
& = - \bigg[ \frac{2}{d} w_1 + \frac{16}{15} w_2 \bigg(\frac{\pi T}{\epsilon_F}\bigg)^2 \bigg] \frac{1}{1 + \tau_{ei}/\tau_{ee}}.
\label{AnalyticCorrection_ThermoelectricPower}
\end{align}
These results show the additional temperature dependence that arises from the momentum-dependent contributions to the elastic scattering rate. These formulas clearly reflect the general features discussed in connection with Eqs.~\eqref{ElectricConductivity_Keyes_1st_Order}-\eqref{ThermoelectricPower_Keyes_1st_Order}.

The simplicity of the RTA in Keyes model gives us the opportunity to isolate the impact of the momentum-dependent electron-impurity scattering in the presence of interactions beyond the perturbation expansion. In particular, we are interested in the relative deviations of the conductivities and the Seebeck coefficient from their values in the presence of strictly momentum-independent elastic scattering [Eqs.~\eqref{ElectricConductivity_0}-\eqref{Thermoelectricpower_0}]
\begin{align}
\frac{\delta A}{A^{(0)}} 
= \frac{A-A^{(0)}}{A^{(0)}}, 
\qquad A\in\{\sigma, \kappa, S\}.
\label{Deviations}
\end{align}

In Fig.~\ref{Figure5}, we present the relative deviations of the electric/thermal conductivities and the Seebeck coefficient for $\delta\Gamma_{\bf p}/\gamma_{ei} \approx w_1 \xi_{\bf p}/\epsilon_F$. For every value of $w_1$, the deviations were found within perturbation theory (dashed lines) and exactly (solid line). The former cases are given by  Eqs.~\eqref{ElectricConductivity_Keyes_1st_Order}-\eqref{ThermoelectricPower_Keyes_1st_Order} and in the displayed temperature range well approximated by Eqs.~\eqref{AnalyticCorrection_ElectricConductivity}-\eqref{AnalyticCorrection_ThermoelectricPower}. For the latter cases, we used Eqs.~\eqref{Sigma_EE}, \eqref{ThermoelectricPower}, and \eqref{ThermalConductivity}. In Figs.~\ref{Figure5}(a)-(c), the relation between $\tau_{out}$ and $\tau_{ee}$ is fixed by the procedure discussed in the previous section.

\begin{figure}[t]
\centering
\includegraphics[width=0.46\textwidth]
{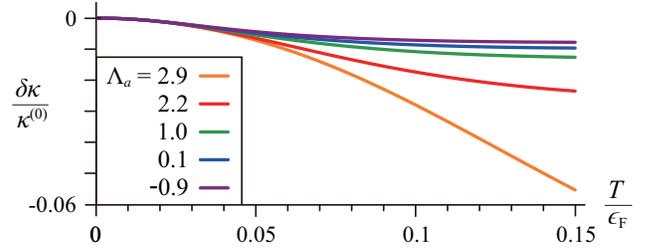} \\
\caption{The relative deviation of the thermal conductivity $\delta\kappa/\kappa^{(0)}$ for FL interactions in three dimensions with various values of  $\Lambda_a$ and fixed  $u = 1.29$ as well as $w_1 = 0.5$.
}
\label{Figure6}
\end{figure}

Overall, the first-order result gives an excellent estimate both of the general trends and the magnitude of the effect. Thus, the analysis via perturbation theory provides us with an accurate picture for the temperature dependence of the transport coefficients. Specifically, we can see the quadratic temperature dependence of the conductivity originating from the averaged energy $\left\langle\!\left\langle \xi_{\bf p}\right\rangle\!\right\rangle\propto T^2/\epsilon_F$ in Fig.~\ref{Figure5}(a). Figure~\ref{Figure5}(c) shows the relative deviation for the Seebeck coefficient. In this case, the $\tau_{ee}$ dependence is induced by $\delta \Gamma_{\bf p}$. The deviation has the maximum strength, estimated by $- w_1$, at zero temperature, but diminishes for higher temeperatures as expected from Eq.~\eqref{AnalyticCorrection_ThermoelectricPower}.

The three-dimensional FL case is shown in Fig.~\ref{Figure6}. Here we observe the strong sensitivity of $\delta\kappa/\kappa^{(0)}$ to the angular dependence of the electron-electron scattering characterized by $\Lambda_a$, similar to that shown in Fig.~\ref{Figure4}(b) for $\kappa^{(0)}$. In particular, changing the dominant collision angles via $\Lambda_a$ for a given temperature amounts to tuning $\tau_{ee}/\tau_{out}$ in the RTA.

\subsubsection{Non-analytic form of the elastic scattering rate}

The momentum dependence of the elastic scattering rate may contain non-analytic contributions. A known example arises due to the modification of the disorder potential by density modulations (Friedel oscillations) forming near impurities in electron liquids. For sufficiently weak disorder, the temperature for which the elastic and inelastic scattering rates are of similar order may be reached for $T\ll \epsilon_F$~\cite{Remark_semiclassical}. Consequently, the $1/\tau_{ee}$ dependence induced by the non-analyticity in $\tau_{ei,{\bf p}}$ could have a strong influence on the temperature dependence of the transport coefficients in the degenerate electron liquid. Our goal here is to describe and illustrate the impact such a term has on various transport coefficients.

To this end, we consider a simple model for the momentum dependent contribution to the elastic scattering rate, $\delta \Gamma_{\bf p}/\gamma_{ei}=\tilde{w}_1\Theta(\xi_{\bf p})\xi_{\bf p}/\epsilon_F$, where $\Theta(x)$ is the Heaviside step function. This model is motivated by the microscopic treatment of electron-impurity scattering in the presence of electron-electron interactions for the two-dimensional electron liquid \cite{Gold86,DasSarma99,Zala01}. In this approach, the simple functional form we use is obtained when the temperature dependence of the Friedel oscillations is neglected. Despite this simplification, the model should be sufficient for capturing the key qualitative consequences of the non-analytic momentum dependence of the scattering rate. Here, we implicitly assume that the inelastic scattering does not substantially modify the Friedel oscillations. The range of these oscillations is of the order of $v_F/T$, while the average distance between elastic scattering events, the mean free path $l_{ei}=v_F\tau_{ei}$, can be much larger in the ballistic regime characterized by the inequalities $1/\tau_{ei}<T<\epsilon_F$. The effects we will discuss here most important for $\tau_{ei}\sim\tau_{ee}$, and therefore inelastic scattering events are unlikely to occur in the area affected by the Friedel oscillations.

The first-order corrections in $\delta \Gamma_{\bf p}$, Eqs.~\eqref{ElectricConductivity_Keyes_1st_Order}-\eqref{ThermoelectricPower_Keyes_1st_Order}, are easily evaluated at low temperatures $(T\ll \epsilon_F)$ for $d=2$, 
\begin{align}
\frac{\delta \sigma^{(1)}}{\sigma_0} 
& = - \ln(2) \tilde{w}_1 \frac{T}{\epsilon_F},
\\
\frac{\delta \kappa^{(1)}}{\kappa^{(0)}}
& = - \frac{27\zeta(3)}{2\pi^2} \tilde{w}_1 \frac{T/\epsilon_F}{1 + \tau_{ei}/\tau_{ee}},
\\
\frac{\delta S^{(1)}}{S^{(0)}}
& = - \frac{\tilde{w}_1}{2(1 + \tau_{ei}/\tau_{ee})},\label{eq:dS_S}
\end{align}
where $\zeta(x)$ is the Riemann zeta function. These corrections show a different temperature dependence compared to the analytic form $\delta \Gamma_{\bf p}\propto \xi_{\bf p}$ considered above.

In Fig.~\ref{Figure7}, we show the relative deviations for the electric/thermal conductivities, and the Seebeck coefficient. Similar to Fig.~\ref{Figure5}, we plot both the perturbative and the exact solutions, and use the same $\tau_{ee}$.  
For $\tilde{w}_1 > 0$, the overall trend is quite similar to Fig.~\ref{Figure5}. At low temperatures, the correction to the electric conductivity features the well-known linear temperature dependence caused by the Friedel oscillations formed around each impurity~\cite{DasSarma99,Gold86,Zala01,Noh03,Kravchenko03}. The dashed line is the linear approximation in $\tilde{w}_1$, which does not depend on $\tau_{ee}$. As this approximation is fairly accurate over the entire range displayed in Fig.~\ref{Figure7}(a), we conclude that the linear temperature dependence we see here is robust against the influence of electron-electron collisions, even beyond the temperature for which $\tau_{ee}=\tau_{ei}$. At sufficiently low temperatures, this correction dominates the $T^2$ dependence obtained from the analytic expansion in the previous section.

\begin{figure}[t]
\centering
\includegraphics[width=0.46\textwidth]
{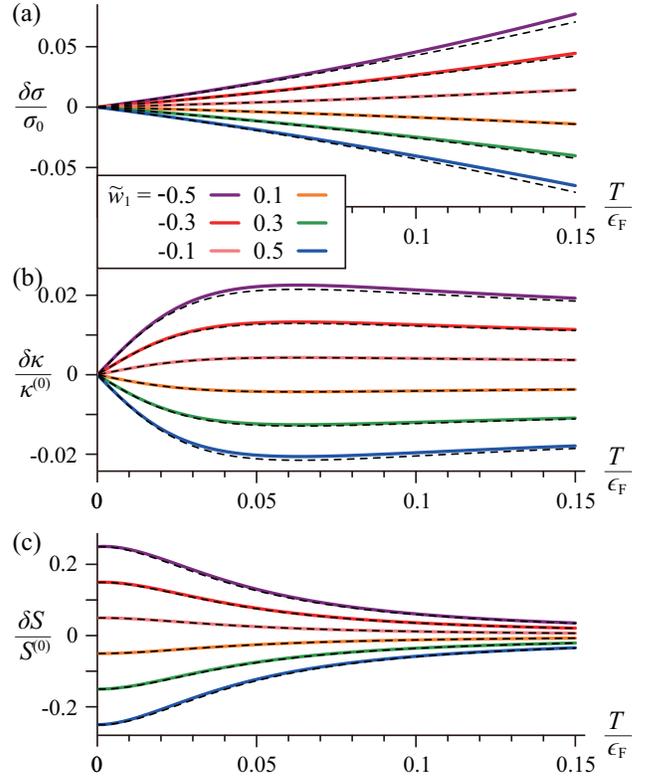} \\
\caption{
The counterpart of Fig.~\ref{Figure5}(a)-(c) for the non-analytic form $\delta\Gamma_{\bf p} = \tilde{w}_1 \gamma_{ei}\Theta(\xi_{\bf p}) \xi_{\bf p}/\epsilon_F$ of the elastic scattering rate.
}
\label{Figure7}
\end{figure}

For $\delta S/S^{(0)}$ as displayed in Fig.~\ref{Figure7}(c), we see a dependence resembling that of the analytic correction shown in Fig.~\ref{Figure5}(c). Indeed, unlike for the electric conductivity, both the analytic term studied in the previous section and the non-analytic term studied here give rise to the same temperature dependence at $T\ll \epsilon_F$. This temperature dependence arises solely from $\tau_{ee}$, and reflects the functional form of the factor $(1+\tau_{ei}/\tau_{ee})^{-1}$ in Eq.~\eqref{eq:dS_S}. As a consequence, in systems where both analytic and non-analytic contributions to $\delta \Gamma_{\bf p}$ exist, they need to be studied on equal footing unless one of them is parametrically small. Figures~\ref{Figure5}(c) and~\ref{Figure7}(c) clearly demonstrate that for $\delta S/S^{(0)}$ the sensitivity to $\tau_{ee}$ induced by the momentum dependence of the elastic scattering rate (both analytical and non-analytical) is crucial for capturing the substantial temperature dependence in the intermediate temperature regime for which $\tau_{ei}\sim\tau_{ee}$. Note that we do not consider here effects of valley degeneracy \cite{Punnoose01} which may considerably increase the strength of the
non-analytic corrections.

\subsection{Non-monotonic behavior and sign change of the Seebeck coefficient}
\label{Sec:Seebeck}

We can make an interesting observation concerning the temperature dependence of the Seebeck coefficient. Focusing on the low-temperature regime ($T\ll \epsilon_F)$ in Eq.~\eqref{ThermoelectricPower_Keyes_1st_Order}, the first term on the right-hand side gives the leading temperature dependence. This term is also the origin of the $w_1$ dependence in the low-temperature expansion given in Eq.~\eqref{AnalyticCorrection_ThermoelectricPower}. As one can immediately see from Eq.~\eqref{AnalyticCorrection_ThermoelectricPower}, $\delta S^{(1)}$ is not necessarily smaller than $S^{(0)}$. Generally speaking, they can be of the same order of magnitude, and correspondingly $w_1\sim 1$, because they both depend on the same effect, the particle-hole asymmetry. Indeed, $S^{(0)}\propto \left\langle\!\left\langle \xi_{\bf p}\right\rangle\!\right\rangle$ is finite either due to the $\xi_{\bf p}$ dependence of $v_{\bf p}^2\propto \xi_{\bf p}+\mu$ entering the momentum integral in the definition of the average, Eq.~\eqref{FluctuationAverage}, or due to the $\xi_{\bf p}$ dependence of the density of states which becomes explicit once the integration variable is changed from ${\bf p}$ to $\xi_{\bf p}$. $\delta S^{(1)}$, in turn, is finite due to the $\xi_{\bf p}$ dependence of the scattering rate encoded in $\delta \Gamma_{\bf p}$. In three dimensions, for example, the origin of this dependence may (again) be the density of states. The origin of the particle-hole asymmetry necessary to render $S$ finite is therefore similar in both cases, and $S^{(0)}$ and $\delta S^{(1)}$ can be of the same order.

For $w_1>0$, a natural behavior in three-dimensional system, when the density of states mainly determines the momentum-dependence of the elastic scattering rate, one finds that $S^{(0)}$ and $\delta S^{(1)}$ have opposite signs. This may lead to an interesting non-monotonic temperature dependence of the Seebeck coefficient if $w_1>d/2$, as illustrated in Fig.~\ref{Figure8} for a three-dimensional system. At the lowest temperatures, $S$ is bound to vanish. At low but finite temperatures, $S$ then takes a positive value since $\delta S^{(1)}$ is dominant. When increasing the temperature further, a change in the sign of $S$ can occur because $\tau_{ee}$ becomes shorter and starts suppressing $\delta S^{(1)}$ relative to $S^{(0)}$; compare Eq.~\eqref{AnalyticCorrection_ThermoelectricPower}. After the sign change occurred when raising the temperature, the Seebeck coefficient is generically negative as can be seen from Eq.~\eqref{Thermoelectricpower_0}. The non-monotonic temperature dependence described above could be utilized in experiment to extract the relative strength of the two scattering mechanisms.

Finally, let us mention that $\delta S^{(1)}$ has an appealing structure, $\delta S^{(1)}\propto \kappa^{(0)}\partial \tau_{ei,{\bf p}}^{-1}/\partial \xi_{\bf p}$, i.e. it resembles the thermal conductivity albeit transformed for the response to the electric field.

\section{Conclusion}
\label{Sec:Conclusion}

We studied the electric, thermal and thermoelectric transport coefficients in the presence of elastic and inelastic collisions. Our approach was based on the Boltzmann equation for analyzing the effect of momentum-dependent collisions. For this purpose, two different solution methods were employed: (i) Keyes approach was generalized to derive the transport coefficients when the elastic scattering rate changes with momentum while the inelastic collision rate is momentum-independent. (ii) An eigenfunction expansion was used for obtaining the conductivities in the case of a constant elastic scattering rate and Fermi liquid and/or Coulomb interactions. Computing the transport coefficients in the various limits provided us with a clear understanding of their sensitivity to electron-electron collisions. We found that inelastic collisions can significantly affect both the thermal conductivity and the Seebeck coefficient. In particular, we showed that the latter can undergo a sign change with increasing inelastic scattering rate or, equivalently, as a function of temperature.   

\begin{figure}[t]
\centering
\includegraphics[width=0.42\textwidth]
{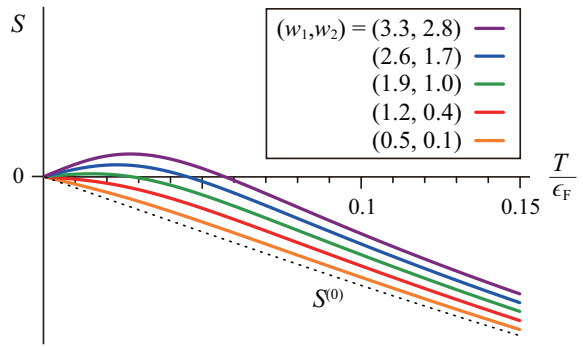} \\
\caption{
The Seebeck coefficient in arbitrary units as a function of $T/\epsilon_F$. All curves were obtained by treading the electron-electron collision integral within the RTA and taking the momentum-dependent part of the electron-impurity scattering rate to be of the form $\delta\Gamma_{\bf p}/\gamma_{ei} = w_1 \xi_{\bf p}/\epsilon_F + w_2 (\xi_{\bf p}/\epsilon_F)^2$. Different values of $w_1$ and $w_2$ lead to the solid lines for fixed $\gamma_{ei}/\epsilon_F = 0.01$ and $u = 1.29$. The dotted black line represents $S^{(0)}$ for the same value of $\gamma_{ei}/\epsilon_F$. A sign change of $S$ occurs for $w_1>1.5$.} 
\label{Figure8}
\end{figure}

One of the main results of the paper is that for Fermi liquid as well as Coulomb interactions the RTA captures the temperature dependence of the transport coefficients well for a constant elastic scattering rate. This observation is particularly important for the thermal conductivity, which is affected by the electron-electron collisions even in this situation. Moreover, it suggest a general guideline for understanding thermal conductivity in other systems: when there are no other energy scales and the temperature and energy enter similarly into the collision integral as $\xi_{\mathbf{p}}^2+\pi^2T^2$ (or, equivalently, into the quasi-particle self-energy), we expect the RTA to be sufficient. Then, $\kappa$ can be found simply by calculating  $1/\tau_{ee}$. By contrast, we expect the collision integral to acquire a more complicated dependence on $\xi_{\mathbf{p}}$ and $T$ when additional energy scales are present.

Finally, it would be very interesting to study the different transport coefficients when elastic and inelastic scattering rates are both momentum-dependent. This would lead to a better understanding of the accuracy of the RTA for inelastic collisions also when the elastic scattering rate depends on momentum. While a general solution of this problem would require the development of a new calculation scheme, we were able to gain insights by perturbatively extending the eigenfunction expansion technique. We found that under the conventional simplifications of Fermi liquid theory, a constant inelastic relaxation rate gives a good approximation for obtaining the thermal and the electric conductivities.

\acknowledgments

We thank A.~Principi, G.~Refael, K.~Tikhonov, and M.~Zarenia for discussions and useful comments. This work was supported by the College of Arts and Sciences at the University of Alabama (W.~L., G.~S.) and the National Science Foundation under Grant No. DMR-1742752 (G.~S.). A.~F.~was supported by the U.S.~Department of Energy, Office of Basic Energy Sciences, Division of Materials Sciences and Engineering under Award DE-SC0014154. K.~M.~was supported by grant No. 2017608 from the United States-Israel Binational Science Foundation (BSF). This work was performed in part at Aspen Center for Physics, which is supported by National Science Foundation grant PHY-1607611 and it was partially supported by a grant from the Simons Foundation.

\appendix

\section{Relaxation-time approximation for $I_{ee}$}
\label{Appendix:RTAforSee}

In this appendix, we provide a more careful justification for the form of the electron-electron collision integral $I_{ee}$ in the RTA as presented in Eq.~\eqref{Seelinear}. For the sake of notational simplicity, we consider a spinless system, and drop the degeneracy factor $s$. For a related discussion in the zero-temperature limit, see Ref.~\cite{Mendl19}.

For the purpose of this discussion, we first present the collision integral in the RTA as
\begin{align}
I_{ee}\{f\} = - \frac{f_{\bf p} - \mathcal{F}_{\rm eq}\{f\}}{\tau_{ee}}.
\end{align}
Here, the dependence of $\mathcal{F}_{\rm eq}$ on $f_{\bf p}$ originates from the $f$ dependence of $\beta$, $\mu$, and ${\boldsymbol v}_{\rm cm}$,
\begin{align}
\mathcal{F}_{\rm eq}\{f\}
= \mathcal{F}_{\rm eq}(\beta\{f\},\mu\{f\}, {\boldsymbol v}_{\rm cm}\{f\}).
\end{align}
A natural choice for $\mathcal{F}_{\rm eq}$ is the drifting Fermi-Dirac distribution
\begin{align}
\mathcal{F}_{\rm eq}\{f\}
= n_F(\epsilon_{\bf p} - \mu\{f\} - {\bf p} \cdot {\boldsymbol v}_{\rm cm}\{f\}),
\label{eq:Feqexp}
\end{align}
where we kept the dependence on $\beta\{f\}$ in the right hand side implicit in order to stay in line with previous notation. The parameters $\beta$, $\mu$, and ${\boldsymbol v}_{\rm cm}$ are determined by the conditions
\begin{align}
\int_{\bf p} \mathcal{F}_{\rm eq}(\beta\{f\},\mu\{f\},{\boldsymbol v}_{\rm cm}\{f\},)
& = \int_{\bf p} f_{\bf p}
= \mathcal{N}, 
\label{eq:N}
\\
\int_{\bf p} {\bf p} \mathcal{F}_{\rm eq}(\beta\{f\},\mu\{f\},{\boldsymbol v}_{\rm cm}\{f\})
& = \int_{\bf p} {\bf p} f_{\bf p}
= \bm{\mathcal{P}},
\\
\int_{\bf p} \epsilon_{\bf p} \mathcal{F}_{\rm eq}(\beta\{f\},\mu\{f\},{\boldsymbol v}_{\rm cm}\{f\})
& = \int_{\bf p}\epsilon_{\bf p} f_{\bf p}
= \mathcal{E}.
\label{eq:E}
\end{align}
Using the form of $\mathcal{F}_{\rm eq}$ given in Eq.~\eqref{eq:Feqexp} and the identity $\epsilon_{\bf p} -  {\bf p} \cdot {\boldsymbol v}_{\rm cm} = \epsilon_{{\bf p} - m{\boldsymbol v}_{\rm cm}} - \frac{1}{2} m {\boldsymbol v}_{\rm cm}^2$, one obtains the relation $\bm{\mathcal{P}}=\mathcal{N}m{\boldsymbol v}_{\rm cm}\{f\}$ and finally an explicit formula for the $f$ dependence of ${\boldsymbol v}_{\rm cm}$, 
\begin{align}
{\boldsymbol v}_{\rm cm}\{f\}
= \frac{\int_{\bf p} {\bf p} f_{\bf p}}{m\int_{\bf p} f_{\bf p}}. 
\label{eq:vf}
\end{align}
At finite temperatures, the definitions for $\mu\{f\}$ and $\beta\{f\}$ remain implicit in Eqs.~\eqref{eq:N} and \eqref{eq:E}.

We now linearize $\mathcal{F}_{\rm eq}$ around the distribution function $f_{\bf p}^{(0)} = \mathcal{F}_{\rm eq}\{f^{(0)}\}$ as
\begin{align}
\mathcal{F}_{\rm eq}\{f^{(0)} + \delta f\}
= f_{\bf p}^{(0)} + \int d{\bf p} \left.\frac{\delta\mathcal{F}_{\rm eq}\{f\}}{\delta f_{\bf p}}\right|_{f_{\bf p} = f_{\bf p}^{(0)}}\delta f_{\bf p},
\label{eq:func}
\end{align}
where we expand using the chain rule
\begin{align}
\frac{\delta \mathcal{F}_{\rm eq}\{f\}}{\delta f_{\bf p}}
& = \frac{\partial \mathcal{F}_{\rm eq}}{\partial \beta}\frac{\delta \beta\{f\}}{\delta f_{\bf p}} + \frac{\partial \mathcal{F}_{\rm eq}}{\partial \mu}\frac{\delta \mu\{f\}}{\delta f_{\bf p}} 
\nonumber\\
& ~~~ + \frac{\partial \mathcal{F}_{\rm eq}}{\partial \boldsymbol{v}_{\rm cm}} \cdot \frac{\delta \boldsymbol{v}_{\rm cm}\{f\}}{\delta f_{\bf p}}.
\label{eq:func1}
\end{align}
In this expression, the derivatives $\delta \mu/\delta f_{\bf p}$ and $\delta \beta/\delta f_{\bf p}$ can be found by differentiating Eqs.~\eqref{eq:N} and \eqref{eq:E}, using the explicit form of $\delta {\boldsymbol v}_{\rm cm}/\delta f_{\bf p} = ({\bf p}-m{\boldsymbol v}_{\rm cm})/((2\pi)^2\mathcal{N}m)$ obtained from Eq.~\eqref{eq:vf}, and solving the resulting system of equations. However, the general form is not very instructive for our purposes. A considerable simplification occurs in the special case ${\boldsymbol v}_{\rm cm}\{f^{(0)}\}=0$ that is relevant to our problem, for which $f^{(0)}_{\bf p} = n_F(\xi_{\bf p})$. We may anticipate further that the solution of the Boltzmann equation will have a form $\delta f_{\bf p}\propto {\bf Y}\cdot{\bf p}$ with the driving field ${\bf Y}\in\{{\bf E},\nabla_{\bf r} T\}$. Since the scalar quantities $(\partial \mathcal{F}_{\rm eq}/\partial \beta) (\delta \beta/\delta f_{\bf p})$ and $(\partial \mathcal{F}_{\rm eq}/\partial \mu) (\delta \mu/\delta f_{\bf p})$ depend only on $|{\bf p}|^2$ for ${\boldsymbol v}_{\rm cm}\{f^{(0)}\}=0$, their contribution to the right-hand side of Eq.~\eqref{eq:func} vanishes upon angular integration. With the explicit form of $\delta {\boldsymbol v}_{\rm cm}/\delta f_{\bf p}$ at hand, one obtains the final form of the linearized collision integral, Eq.~\eqref{Seelinear} in the main text, with ${\boldsymbol v}_{\rm cm} = \int_{\bf p} {\bf p}\delta f_{\bf p} / (\mathcal{N}m)$.

\section{Derivation of the approximate form of the electron-electron collision integral}
\label{Appendix_Approximate_e-e_Coll}

\subsection{FL-type collision integral in three dimensions}

In this section, we derive the approximate form of the collision integral given in Eq.~\eqref{Coll_Int_ee_3} for the case of a FL in three dimensions ($d=3$)~\cite{Abrikosov59}. Using Eq.~\eqref{Coll_Int_ee_2} as a starting point, we first introduce $\omega$ as the energy transferred during the electron-electron scattering by using the identity $\delta(\epsilon_{\bf p} + \epsilon_{\bf q} - \epsilon_{{\bf p}'} - \epsilon_{{\bf q}'}) = \int_{-\infty}^\infty d\omega \delta(\omega - \epsilon_{{\bf p}'} + \epsilon_{\bf p}) \delta(\omega - \epsilon_{\bf q} + \epsilon_{{\bf q}'})$. This allows us to write the collision integral as
\begin{align}
&I_{ee}\{\phi\} 
= - \frac{2(2\pi)^4}{T} \sum_{\alpha\in \{E,T\}} \int_{{\bf q},{\bf p'},{\bf q'}} \int_{-\infty}^\infty d\omega |U_{{\bf p}{\bf q},{\bf p}'{\bf q}'}|^2
\nonumber\\
&~~~~~ \times \delta (\omega-\epsilon_{\bf p'}+\epsilon_{\bf p}) \delta(\omega-\epsilon_{\bf q}+\epsilon_{\bf q'})\delta ({\bf p}+{\bf q}-{\bf p}'-{\bf q}')
\nonumber\\
&~~~~~ \times n_F(\xi_{\bf p}) n_F(\xi_{\bf q}) [1 - n_F(\xi_{\bf p} + \omega)] [1 - n_F(\xi_{\bf q} - \omega)] 
\nonumber\\
&~~~~~ \times [\phi^{\alpha}(\xi_{\bf p}){\boldsymbol{v}}_{\bf p}+\phi^{\alpha}(\xi_{\bf q})\boldsymbol{v}_{\bf q} -\phi^{\alpha}(\xi_{\bf p}+\omega)\boldsymbol{v}_{\bf p'}
\nonumber \\
&~~~~~ - \phi^{\alpha}(\xi_{\bf q}- \omega)\boldsymbol{v}_{\bf q'}] \cdot {\bf F}_\alpha,
\end{align}
where ${\bf F}_E = - e {\bf E}$ and ${\bf F}_T = -\nabla_{\bf r} T$. We next separate the angular part of the momentum integrals by writing $\int_{\bf q} = \int d\hat{\bf n}_{\bf q} \int_{-\epsilon_F}^\infty d\xi_{\bf q} \nu_3(\epsilon_{\bf q})$, where we defined the unit vector $\hat{\bf n}_{\bf q} = {\bf q} / |{\bf q}|$, used the normalization $\int d\hat{\bf n}_{\bf q} = 1$, and introduced the density of states in three dimensions $\nu_{3}(\epsilon_{\bf q})=(2m)^{3/2} \sqrt{\epsilon_{\bf q}} / (4\pi^2)$.

The integrations in $\xi_{\bf p'}$ and $\xi_{\bf q'}$ are trivially performed with the help of the delta functions containing $\omega$. With the aim of extracting the leading dependence on $T$ and $\xi_{\bf p}$, we may project the density of states, the velocities and the interaction matrix element onto the Fermi surface, and approximate $\delta({\bf p}+{\bf q}-{\bf p'}-{\bf q'})\approx p^{-3}_F \delta(\hat{\bf n}_{\bf p} + \hat{\bf n}_{\bf q} - \hat{\bf n}_{\bf p'} - \hat{\bf n}_{\bf q'})$. These steps result in 
\begin{align}
I_{ee}\{\phi\}
& = - \frac{2(2\pi)^4}{T} [\nu_{3}(\epsilon_F)]^3 \sum_{\alpha\in \{E,T\}}\frac{F_\alpha v_F}{p_F^3} \int_{-\epsilon_F}^\infty d\xi_{\bf q} 
\nonumber\\
& \times \int_{-\infty}^\infty d\omega n_F(\xi_{\bf p}) n_F(\xi_{\bf q}) [1 - n_F(\xi_{\bf p} + \omega)] 
\nonumber\\
& \times [1 - n_F(\xi_{\bf q} - \omega)] {\bf \Psi}_\alpha(\hat{\bf n}_{\bf p};\xi_{\bf p},\xi_{\bf q},\omega) \cdot \hat{{\bf n}}_{{\bf F}_{\alpha}},
\end{align}
where we wrote ${\bf F}_\alpha = F_\alpha \hat{\bf n}_{{\bf F}_{\alpha}}$, and introduced
\begin{align}
& {\bf \Psi}_\alpha(\hat{{\bf n}}_{\bf p};\xi_{\bf p},\xi_{\bf q},\omega)
= \int d\hat{{\bf n}}_{\bf q} d\hat{{\bf n}}_{\bf p'}d\hat{{\bf n}}_{\bf q'} |\tilde{U}_{\hat{{\bf n}}_{\bf p}\hat{{\bf n}}_{\bf q}\hat{{\bf n}}_{\bf p'}\hat{{\bf n}}_{\bf q'}}|^2
\nonumber\\
& ~~~ \times \delta(\hat{{\bf n}}_{\bf p}+\hat{{\bf n}}_{\bf q}-\hat{{\bf n}}_{\bf p'}-\hat{{\bf n}}_{\bf q'}) [\phi^\alpha({\xi}_{\bf p})\hat{{\bf n}}_{\bf p} + \phi^\alpha({\xi}_{\bf q})\hat{{\bf n}}_{\bf q} 
\nonumber\\
& ~~~ - \phi^\alpha(\xi_{\bf p} + \omega)\hat{{\bf n}}_{\bf p'} - \phi^\alpha(\xi_{\bf q} - \omega) \hat{{\bf n}}_{\bf q'}],
\end{align}
where $\tilde{U}$ is obtained from $U$ by fixing all momenta to $p_F$. For parametrization of the matrix element $\tilde{U}$, it is customary in FL theory to use two angles: $\theta$ is the angle between the two incoming momenta ${\bf p}$ and ${\bf q}$, and $\varphi$ is the angle between the planes spanned by ${\bf p}$ and ${\bf q}$ and by ${\bf p'}$ and ${\bf q'}$, respectively. With this parametrization, the momentum transfer is $k = |{\bf p}-{\bf p}'| = 2p_F \sin(\theta/2) \cos(\varphi/2)$.

We focus on the angular integral ${\mathbf \Psi}_\alpha$ now. The integration in $\hat{{\bf n}}_{\bf p'}$ and $\hat{{\bf n}}_{\bf q'}$ is most conveniently performed in spherical coordinates with the $z$ axis aligned with the total momentum of the incoming particles, i.e., along $\hat{{\bf n}}_{\bf p}+\hat{{\bf n}}_{\bf q}=\hat{{\bf n}}_{\bf p'}+\hat{{\bf n}}_{\bf q'}$, with the help of the identity
\begin{align}
& ~~~ \delta(\hat{{\bf n}}_{\bf p}+\hat{{\bf n}}_{\bf q}-\hat{{\bf n}}_{\bf p'}-\hat{{\bf n}}_{\bf q'})
\nonumber\\
& = \frac{\delta(\theta_{\bf p'}-\theta_{\bf q'})\delta(\varphi_{\bf q'}-\varphi_{\bf p'}-\pi)\delta(\theta_{\bf p'}-\theta/2)}{2\cos(\theta/2) \sin^2(\theta/2)}.
\end{align}
We see that the angle between the outgoing momenta equals that between the incoming momenta, $\theta$. We can adjust the coordinate system so that $\varphi_{{\bf p}'} = \varphi$. It is important to note that the coordinate system used so far depends on $\hat{\bf n}_{\bf q}$. We now choose coordinates so that the $z$-axis points along $\hat{\bf n}_{\bf p}$, and the polar angle coincides with the angle $\theta$ used for the parametrization of the interaction matrix element, $\theta_{\bf q}=\theta$. The unit vectors $\hat{\bf n}_{\bf q}$, $\hat{\bf n}_{\bf p'}$ and $\hat{\bf n}_{\bf q'}$ depend on $\varphi_{\bf q}$, and need to be averaged. Denoting $\bar{\bf n}_{\bf k} = \int d(\varphi_{\bf q}/2\pi)\hat{{\bf n}}_{\bf k}$, one finds
\begin{align}
{\mathbf \Psi}_\alpha(\hat{{\bf n}}_{\bf p};\xi_{\bf p},\xi_{\bf q},\omega)
& = \frac{\pi}{(4\pi)^3} \int_0^\pi \frac{\sin\theta d\theta}{\cos(\theta/2)} \int_0^{2\pi} d\varphi |\tilde{U}(\theta,\varphi)|^2
\nonumber\\
& \times [\phi^\alpha({\xi}_{\bf p})\bar{\bf n}_{\bf p} + \phi^\alpha({\xi}_{\bf q})\bar{\bf n}_{\bf q} - \phi^\alpha(\xi_{\bf p} + \omega)\bar{\bf n}_{\bf p'}
\nonumber\\
& - \phi^\alpha(\xi_{\bf q} - \omega) \bar{\bf n}_{\bf q'}],
\label{eq:Psi}
\end{align}
with $\bar{\bf n}_{\bf p} = \hat{\bf n}_{\bf p}$, $\bar{\bf n}_{\bf q} = \cos\theta~\hat{\bf n}_{\bf p}$, and $\bar{\bf n}_{{\bf p'}/{\bf q'}} = [\cos^2\theta \mp \cos\varphi$ $\sin^2(\theta/2)] \hat{\bf n}_{\bf p}$. Returning to the full collision integral, the integrations in $\omega$ and $\xi_{\bf q}$ remain, where the lower integration limit is safely extended to $-\infty$. For this, we use
\begin{align}
& ~~~ \int_{-\infty}^\infty d\xi_{\bf q} \int_{-\infty}^\infty d\omega n_F(\xi_{\bf q}) [1 - n_F(\xi_{\bf p} + \omega)] 
\nonumber\\
& ~~~ \times [1 - n_F(\xi_{\bf q} - \omega)] \phi^{\alpha}(x_j)
\nonumber\\
& = \int d\omega \omega n_B(\omega) [1 - n_{F}(\xi_{\bf p}+\omega)] \phi^{\alpha}(y_j),
\label{eq:enint}
\end{align}
for $(x_1, x_2, x_3, x_4) = (\xi_{\bf p}, \xi_{\bf q}, \xi_{\bf p} + \omega, \xi_{\bf q} - \omega)$ and $(y_1, y_2,$ $y_3, y_4) = (\xi_{\bf p}, - \xi_{\bf p} - \omega, \xi_{\bf p} + \omega, \xi_{\bf p} + \omega)$.
In order to obtain these results, we used the identities $n_F(-\xi) = 1 - n_F(\xi)$ and $\int d\xi n_F(\xi) [1 - n_F(\xi-\omega)] = \omega n_B(\omega)$, and shifted or relabeled integration variables when convenient. Combining the results for the angular integral ${\mathbf \Psi}_\alpha$ in Eq.~\eqref{eq:Psi} with those for the energy integrations listed in Eq.~\eqref{eq:enint}, one arrives at the form for collision integral reported in the main text, Eq.~\eqref{Coll_Int_ee_3}.

\subsection{Coulomb collision integral in two dimensions}

In the case of two dimensions ($d = 2$), we consider electron-electron scattering mediated by the Coulomb interaction along similar lines as Ref.~\cite{Lyakhov03}. In the random-phase approximation, the Coulomb interaction has the form
\begin{align}
U({\bf k}, \omega) 
= \frac{U_0(|{\bf k}|)}{1 - s U_0(|{\bf k}|) \chi({\bf k}, \omega)},
\label{CoulombInteraction_1}
\end{align}
where we define the bare Coulomb interaction $U_0(|{\bf k}|) = 2\pi e^2 / k$, and the Lindhard function
\begin{align}
\chi({\bf k}, \omega) 
= \int_{\bf p'} \frac{n_F(\xi_{\bf p'}) - n_F(\xi_{{\bf p'} + {\bf k}})}{\omega + \epsilon_{\bf p'} - \epsilon_{{\bf p'} + {\bf k}} + i\eta},
\label{LindhardFunction}
\end{align}
with $s=2$ counting spin degeneracy, and $\eta\rightarrow0^+$.

We now simplify Eq.~\eqref{Coll_Int_ee_2} step by step. Using the two identities $\delta({\bf p} + {\bf q} - {\bf p}' - {\bf q}') = \int d{\bf k} \delta({\bf k} - {\bf p}' + {\bf p}) \delta({\bf k} - {\bf q} + {\bf q}')$ and $\delta(\epsilon_{\bf p} + \epsilon_{\bf q} - \epsilon_{{\bf p}'} - \epsilon_{{\bf q}'}) = \int d\omega \delta(\omega - \epsilon_{{\bf p}'} + \epsilon_{\bf p}) \delta(\omega - \epsilon_{\bf q} + \epsilon_{{\bf q}'})$ with momentum transfer ${\bf k}$ and energy transfer $\omega$, Eq.~\eqref{Coll_Int_ee_2} turns into
\begin{align}
I_{ee}\{\phi\} 
& = - \frac{4\pi}{mT} \sum_{\alpha\in\{E,T\}} \int_{\bf q,k} \int_{-\infty}^\infty d\omega |U(\epsilon_{\bf k}, \omega)|^2 
\nonumber\\
& \times \delta(\omega - \epsilon_{\bf p+k} + \epsilon_{\bf p}) \delta(\omega - \epsilon_{\bf q} + \epsilon_{\bf q-k}) 
\nonumber\\
& \times n_F(\xi_{\bf p}) n_F(\xi_{\bf q}) [1 - n_F(\xi_{\bf p} + \omega)] [1 - n_F(\xi_{\bf q} - \omega)] 
\nonumber\\
& \times \big[ \{\phi^\alpha(\xi_{\bf p}) - \phi^\alpha(\xi_{\bf p} + \omega)\} {\bf p} 
\nonumber\\
& + \{\phi^\alpha(\xi_{\bf q}) - \phi^\alpha(\xi_{\bf q} - \omega)\} {\bf q} 
\nonumber\\
& + \{\phi^\alpha(\xi_{\bf q} - \omega) - \phi^\alpha(\xi_{\bf p} + \omega)\} {\bf k} \big] \cdot {\bf F}_\alpha.
\label{Coll_Int_CI_1}
\end{align}
If we separate the angular part from the coordinate such that $\int_{\bf q} = \int_{0}^{2\pi} d(\theta_{\bf q}/2\pi) \int_{0}^{\infty} d\epsilon_{\bf q} \nu_2$ with the angle $\theta_{\bf q}$ defined relative to the reference vector ${\bf F}_\alpha$, and the density of states in two dimensions $\nu_2 = m/(2\pi)$, Eq.~\eqref{Coll_Int_CI_1} is rearranged into
\begin{align}
& I_{ee}\{\phi\} 
= - \frac{4\pi m}{(2\pi)^{4}T} 
\nonumber\\
& \times \sum_{\alpha\in\{E,T\}} F_\alpha \int_{0}^{\infty} d\epsilon_{\bf q} 
\int_{0}^{\infty} d\epsilon_{\bf k} \int_{-\infty}^{\infty} d\omega |U({\bf k}, \omega)|^2 
\nonumber\\
& \times n_F(\xi_{\bf p}) n_F(\xi_{\bf q}) [1 - n_F(\xi_{\bf p} + \omega)] [1 - n_F(\xi_{\bf q} - \omega)] 
\nonumber\\
& \times \big[ \{\phi^\alpha(\xi_{\bf p}) - \phi^\alpha(\xi_{\bf p} + \omega)\} \Xi_1(\theta_{\bf p};\epsilon_{\bf p},\epsilon_{\bf q},\epsilon_{\bf k},\omega)
\nonumber\\
& + \{\phi^\alpha(\xi_{\bf q}) - \phi^\alpha(\xi_{\bf q} - \omega)\} \Xi_2(\theta_{\bf p};\epsilon_{\bf p},\epsilon_{\bf q},\epsilon_{\bf k},\omega)
\nonumber\\
& + \{\phi^\alpha(\xi_{\bf q} - \omega) - \phi^\alpha(\xi_{\bf p} + \omega)\} \Xi_3(\theta_{\bf p};\epsilon_{\bf p},\epsilon_{\bf q},\epsilon_{\bf k},\omega) \big],
\label{Coll_Int_CI_2}
\end{align}
where we define the angular integral parts ($j=1,2,3$) 
\begin{align}
& ~~~  \Xi_j(\theta_{\bf p};\epsilon_{\bf p},\epsilon_{\bf q},\epsilon_{\bf k},\omega)
\nonumber\\
& = \sqrt{2m\epsilon_j} \int_{0}^{2\pi} d\theta_{\bf k} \int_{0}^{2\pi} d\theta_{\bf q} \cos\theta_j 
\nonumber\\
& ~~~ \times \delta\big(\omega - \epsilon_{\bf k} - 2 \sqrt{\epsilon_{\bf p} \epsilon_{\bf k}} \cos(\theta_{\bf k} - \theta_{\bf p}) \big) 
\nonumber\\
& ~~~ \times \delta\big(\omega + \epsilon_{\bf k} - 2 \sqrt{\epsilon_{\bf q} \epsilon_{\bf k}} \cos(\theta_{\bf q} - \theta_{\bf k}) \big),
\label{AngularIntegralCoulomb_1}
\end{align}
with $(\epsilon_1,\theta_1) = (\epsilon_{\bf p},\theta_{\bf p})$, $(\epsilon_2,\theta_2) = (\epsilon_{\bf q},\theta_{\bf q})$, and $(\epsilon_3,\theta_3) = (\epsilon_{\bf k},\theta_{\bf k})$. The result of the integration in \eqref{AngularIntegralCoulomb_1} can be written in the form
\begin{align}
& ~~~  \Xi_j(\theta_{\bf p};\epsilon_{\bf p},\epsilon_{\bf q},\epsilon_{\bf k},\omega)
\nonumber\\
& = \frac{|{\bf p}| \cos\theta_{\bf p} \Pi_j(\epsilon_{\bf p},\epsilon_{\bf k},\omega)}{\sqrt{\epsilon_{\bf p} \epsilon_{\bf k} - [(\omega - \epsilon_{\bf k})/2]^2} \sqrt{\epsilon_{\bf q} \epsilon_{\bf k} - [(\omega + \epsilon_{\bf k})/2]^2}},
\label{AngularIntegralCoulomb_2}
\end{align}
where $\Pi_1 = 1$, $\Pi_2 = (\omega^2 - \epsilon_{\bf k}^2)/(4\epsilon_{\bf p}\epsilon_{\bf k})$, and $\Pi_3 = (\omega - \epsilon_{\bf k})/(2\epsilon_{\bf p})$.

It is convenient to introduce two approximations to proceed. First, in the degenerate regime of $T \ll \epsilon_F$, the combination of distribution functions in Eq.~\eqref{Coll_Int_CI_2} pins all momenta of incoming particles to the Fermi surface such that $\epsilon_{\bf p} \approx \epsilon_{\bf q} \approx \epsilon_F$. Second, in the regime of dominant forward scattering, the phase space factor in Eq.~\eqref{AngularIntegralCoulomb_2} is greatly simplified. We notice that the phase space factor is maximized in case of $(\omega \pm \epsilon_{\bf k})^2 \ll 4\epsilon_F \epsilon_{\bf k}$, which is consistent with $|\omega| \ll 4\epsilon_F$ and $\omega^2 / (4\epsilon_F) \ll \epsilon_{\bf k} \ll 4\epsilon_F$. The scattering angles are governed by the conservation law in Eq.~\eqref{AngularIntegralCoulomb_1} such that ${\bf p} \perp {\bf k}$ and ${\bf q} \perp {\bf k}$, thus we conclude ${\bf p} \parallel {\bf p}'$ and ${\bf q} \parallel {\bf q}'$. We can now simplify
\begin{align}
\Xi_1(\theta_{\bf p};\epsilon_{\bf p},\epsilon_{\bf q},\epsilon_{\bf k},\omega)
\approx \frac{|{\bf p}| \cos\theta_{\bf p}}{\epsilon_F\epsilon_{\bf k}},
\label{AngularIntegralCoulomb_3}
\end{align}
while $\Xi_2 \approx \Xi_3 \approx 0$ are parametrically smaller . In the same parameter regime we may use the universal limit for the Coulomb interaction, 
\begin{align}
U({\bf k}, \omega)\approx \frac{1}{2\chi({\bf k},\omega)},
\label{CoulombInteraction_3}
\end{align}
and further simplify $U({\bf k},\omega)\approx 1/(2\nu_2) = \pi/m$. For $T\ll \epsilon_F$, the remaining energy integral may now be approximated as
\begin{align}
\int_{\omega^2/(4\epsilon_F)}^{4\epsilon_F} \frac{d\epsilon_{\bf k}}{\epsilon_{\bf k}}
\approx 2 \ln \bigg|\frac{4\epsilon_F}{T}\bigg|,
\label{Evaluation_Coulomb}
\end{align}
where we replaced $\omega$ by the typical scale $T$. Collecting all results, we finally derive the approximate form of the electron-electron collision integral, Eq.~\eqref{Coll_Int_ee_3}.

\section{Solving the Boltzmann equation in the FL approximation}
\label{Appendix:solvingFLKE}

Plugging Eq.~\eqref{Coll_Int_ee_3} and $I_{ei}=-\delta f/\tau_{ei}$ with $\delta f$ given in Eq.~\eqref{DistributionFluctuation} in Eq.~\eqref{Coll_Int_ee_2}, taking ${\bf E}$ and $\nabla_{\bf r} T$ as independent, and matching symmetries on both sides, one can represent  the nontrivial parts of the resulting Boltzmann equation in the two-component form
\begin{align}
\left(
\begin{array}{c}
1
\\
\beta\xi_{\bf p}
\end{array}
\right)
& = \frac{1}{\tau_{ei}} \left(
\begin{array}{c}
\phi_s^E(\xi_{\bf p})
\\
\phi_a^T(\xi_{\bf p})
\end{array}
\right) 
\nonumber\\
& ~~~ + \frac{4}{(\pi T)^2 \tau_{out}} \int_{-\infty}^{\infty} d\omega K(\omega,\xi_{\bf p}) 
\nonumber\\
& ~~~ \times \left(
\begin{array}{c}
\phi_s^E(\xi_{\bf p}) - \Lambda_s \phi_s^E(\xi_{\bf p} + \omega)
\\
\phi_a^T(\xi_{\bf p}) - \Lambda_a \phi_a^T(\xi_{\bf p} + \omega)
\end{array}
\right).
\label{BoltzmannEq_Linearized_Refined_2}
\end{align}
Here, it can be shown that $\phi_a^E$ and $\phi_s^T$ trivially vanish. We conveniently recast Eq.~\eqref{BoltzmannEq_Linearized_Refined_2} in the standard form~\cite{Jensen68,Sykes70}:
\begin{align}
\mathbb{X}(x)
& = (x^2 + \pi^2 \varepsilon^2) \mathbb{Q}(x) 
\nonumber\\
& ~~~ - \int_{-\infty}^{\infty} dy (y-x)~{\rm csch}\bigg(\frac{y-x}{2}\bigg) \mathbb{M} \mathbb{Q}(y),
\label{BoltzmannEq_Linearized_Refined_3}
\end{align}
where we define the two-component functions
\begin{align}
\mathbb{X}(x) 
& = \left(
\begin{array}{c}
1
\\
x
\end{array}
\right) {\rm sech}\bigg(\frac{x}{2}\bigg),
\label{Function_X}
\\
\mathbb{Q}(x) 
& = \left(
\begin{array}{c}
\hat{Q}_s^E(x)
\\
\hat{Q}_a^T(x)
\end{array}
\right)
= \frac{2}{\pi^2} \frac{1}{\tau_{out}} \left(
\begin{array}{c}
\hat{\phi}_s^E(x)
\\
\hat{\phi}_a^T(x)
\end{array}
\right) {\rm sech}\bigg(\frac{x}{2}\bigg),
\label{Function_Q}
\end{align}
with respect to the dimensionless variables $x = \beta\xi_{\bf p}$ and $y = \beta\omega$, and the diagonal matrix $\mathbb{M}=\mbox{diag}(\Lambda_s,\Lambda_a)$, 
and introduce the parameter $
\varepsilon = \sqrt{1 + {\tau_{out}}/{2\tau_{ei}}}$. Hereafter, we conveniently define $\hat{Y}(x) = Y(x/\beta) = Y(\xi_{\bf p})$ for $Y\in\{Q,\phi\}$. After a Fourier transformation 
\begin{align}
\tilde{\mathbb{Q}}(\tilde{x}) 
& = \int_{-\infty}^\infty dx e^{i\tilde{x}x} \mathbb{Q}(x), 
\label{FourierTransform_Q}
\end{align}
Eq.~\eqref{BoltzmannEq_Linearized_Refined_3} can be converted to a second-order inhomogeneous differential equation
\begin{align}
\bigg[ \mathbb{I} \frac{d^2}{d\tilde{x}^2} + \pi^2 \big[ 2~{\rm sech}^2(\pi \tilde{x}) \mathbb{M} - \varepsilon^2 \mathbb{I} \big] \bigg] \tilde{\mathbb{Q}}(\tilde{x}) 
= - \tilde{\mathbb{X}}(\tilde{x}),
\label{BoltzmannEq_Linearized_Refined_4}
\end{align}
where $
\tilde{\mathbb{X}}(\tilde{x})= 2\pi~{\rm sech}(\pi \tilde{x}) (
1,i\pi~{\rm tanh}(\pi \tilde{x}))^t$, and $\mathbb{I}$ is the identity matrix. It would be the standard solution strategy for Eq.~\eqref{BoltzmannEq_Linearized_Refined_4} (i) to derive the eigensolutions for the homogeneous equation, and (ii) to find the series solution for the inhomogeneous equation as an eigenfunction expansion.

We first consider the homogeneous equation with $\tilde{\mathbb{X}} = 0$ in Eq.~\eqref{BoltzmannEq_Linearized_Refined_4}, and promote it to the eigenvalue equation with the parameter $\lambda$
\begin{align}
\bigg[ \frac{d^2}{d\tilde{x}^2} + \pi^2 \big[ 2\lambda~{\rm sech}^2(\pi \tilde{x}) - \varepsilon^2 \big] \bigg] \tilde{Q}(\tilde{x}) = 0,
\label{KineticEq_NoMomentum_Homogeneous}
\end{align}
which is analogous to the time-independent Schr\"{o}dinger equation~\cite{LL:QM}. Changing the variable by $\xi = {\rm tanh}(\pi \tilde{x})$, we recast Eq.~\eqref{KineticEq_NoMomentum_Homogeneous} in the self-adjoint form
\begin{align}
\bigg[ (1 - \xi^2) \frac{d^2}{d\xi^2} - 2 \xi \frac{d}{d\xi} + 2\lambda - \frac{\varepsilon^2}{1 - \xi^2} \bigg] \tilde{Q}(\xi) = 0.
\label{AssociatedLegendreDifferentialEq}
\end{align}
In the clean limit of $\varepsilon = 1$, Eq.~\eqref{AssociatedLegendreDifferentialEq} is satisfied by the associated Legendre polynomial $P_{n+1}^{\varepsilon=1}(\xi)$ with $\lambda_n = (n+1)(n+2)/2$ for $n \in \{0, \mathbb{N}\}$. For a disordered system with non-integer $\varepsilon>1$, we need to recast further Eq.~\eqref{AssociatedLegendreDifferentialEq} in terms of $\mathcal{Q}(\xi) = (1 - \xi^2)^{-\varepsilon/2} \tilde{Q}(\xi)$ and $\zeta = (1 - \xi)/2$. For the choice of the eigenvalue $
\lambda_n(\varepsilon) = (n + \varepsilon) (n + \varepsilon + 1)/2$, 
$\mathcal{Q}$ is governed by
\begin{align}
& \bigg[ \zeta (1 - \zeta) \frac{d^2}{d\zeta^2} + (\varepsilon + 1) (1 - 2\zeta) \frac{d}{d\zeta} + n (n + 2\varepsilon + 1) \bigg] 
\nonumber\\
& \times \mathcal{Q}(\zeta) = 0.
\label{HypergeometricDifferentialEq}
\end{align}
The solution is the hypergeometric function $_2 F_1(-n, n + 2\varepsilon + 1, \varepsilon + 1, \zeta)$. After representing the eigenfunction in the original variable $\tilde{x}$, one arrives at  
\begin{align}
\tilde{Q}_n(\tilde{x})
& = [{\rm sech}(\pi \tilde{x})]^\varepsilon~_2 F_1[-n, n + 2\varepsilon + 1, \varepsilon + 1, 
\nonumber\\
& ~~~~ (1 - {\rm tanh}(\pi \tilde{x}))/2]. 
\label{Eigenfunction_NoMomentum}
\end{align}
where $_2 F_1(-n, n + 2\varepsilon + 1, \varepsilon + 1, \zeta)$ is the hypergeometric function.

The eigenfunctions are even(odd)-symmetric in $\tilde{x}$ for even (odd) integer $n$, and satisfy the orthogonal relation
\begin{align}
& \int_{-\infty}^\infty d\tilde{x}~{\rm sech}^2(\pi \tilde{x}) [\tilde{Q}_{m}(\tilde{x})]^* \tilde{Q}_{n}(\tilde{x})
\nonumber\\
& = \frac{n! 2^{2\varepsilon + 1} [\Gamma(\varepsilon + 1)]^2 \delta_{mn}}{\pi (2n + 2\varepsilon + 1) \Gamma(n + 2\varepsilon + 1)},
\label{Orthogonality}
\end{align}
where $\Gamma(z)$ is the Gamma function, and $\delta_{mn}$ is the Kronecker delta.

The series expansion of  $\tilde{\mathbb{Q}}(\tilde{x})$ in the eigenfunctions reads
\begin{align}
\tilde{Q}_s^E(\tilde{x})
& = \sum_{n=0}^\infty C_{2n}^E \tilde{Q}_{2n}(\tilde{x}),
\label{SeriesSolution_E}
\\
\tilde{Q}_a^T(\tilde{x})
& = \sum_{n=0}^\infty C_{2n+1}^T \tilde{Q}_{2n+1}(\tilde{x}),
\label{SeriesSolution_T}
\end{align}
with the expansion coefficients \cite{Bennett69}
\begin{align}
C_{2n}^E
& = \frac{\int_{-\infty}^\infty d\tilde{x}~{\rm sech}(\pi \tilde{x}) [\tilde{Q}_{2n}(\tilde{x})]^*}{\pi[\lambda_{2n}(\varepsilon) - \Lambda_s] \int_{-\infty}^\infty d\tilde{x}~{\rm sech}^2(\pi \tilde{x}) |\tilde{Q}_{2n}(\tilde{x})|^2},
\label{ExpansionCoefficients_E}
\\
C_{2n+1}^T
& = \frac{i \int_{-\infty}^\infty d\tilde{x}~{\rm sech}(\pi \tilde{x}) {\rm tanh}(\pi \tilde{x}) [\tilde{Q}_{2n+1}(\tilde{x})]^*}{[\lambda_{2n+1}(\varepsilon) - \Lambda_a] \int_{-\infty}^\infty d\tilde{x}~{\rm sech}^2(\pi \tilde{x}) |\tilde{Q}_{2n+1}(\tilde{x})|^2},
\label{ExpansionCoefficients_T}
\end{align}
where $\lambda_n(\varepsilon) = {(n + \varepsilon) (n + \varepsilon + 1)}/{2}$. We found that the expansion coefficients can be simplified as
\begin{align}
C_{2n}^E
& = \frac{2n + \varepsilon + 1/2}{\pi [\lambda_{2n}(\varepsilon) - \Lambda_s]} \frac{\Gamma(n+\varepsilon+1/2)\Gamma(n+(\varepsilon+1)/2)}{\Gamma(n+\varepsilon/2+1)\Gamma(n+1)\Gamma(\varepsilon+1)},
\label{Coefficient_E}
\\
C_{2n+1}^T
& = \frac{i(2n + \varepsilon + 3/2)}{\lambda_{2n+1}(\varepsilon) - \Lambda_a} \frac{\Gamma(n+\varepsilon+3/2)\Gamma(n+(\varepsilon+1)/2)}{\Gamma(n+\varepsilon/2+2)\Gamma(n+1)\Gamma(\varepsilon+1)},
\label{Coefficient_T}
\end{align}
In the clean limit of $\varepsilon = 1$, the coefficients reduce to $C_{2n}^E = (4n+3)/[2\pi(\lambda_{2n} - \Lambda_s)]$ and $C_{2n+1}^T = i (4n + 5) / [2(\lambda_{2n+1} - \Lambda_a)]$, which are consistent with Ref.~\cite{Sykes70}.

The above solutions are represented in the conjugate variable $\tilde{x}$, and are useful to determine the linear conductivities. If we want to investigate directly the non-equilibrium distribution fluctuations, we should represent them in the original variable $x = \beta\xi_{\bf p}$, 
\begin{align}
\left(
\begin{array}{c}
\phi_s^E(\xi_{\bf p})
\\
\phi_a^T(\xi_{\bf p})
\end{array}
\right)
& = \frac{\pi^2}{2} \tau_{out} {\rm cosh}\bigg(\frac{\beta\xi_{\bf p}}{2}\bigg) 
\sum_{n=0}^\infty 
\left(
\begin{array}{c}
C_{2n}^E Q_{2n}(\xi_{\bf p})
\\
C_{2n+1}^T Q_{2n+1}(\xi_{\bf p})
\end{array}
\right),
\label{NoneqDistFluc_Refined}
\end{align}
where we define the inverse Fourier transform $\hat{Q}_n(x) = \int_{-\infty}^\infty d(\tilde{x}/2\pi) e^{-i\tilde{x}x} \tilde{Q}_n(\tilde{x})$ that can be written in the closed form given in Eq.~\eqref{InverseFourierTransformQn} of the main text; see Appendix~\ref{Appendix_InverseFourierTransform}.

The thermal and electric conductivities can be obtained from the solution of the Boltzmann equation in the dimensionless variable $x = \beta\xi_{\bf p}$ as
\begin{align}
\left(
\begin{array}{c}
L_{EE}
\\ L_{TT}
\end{array}
\right)
& = \frac{\pi^2\mathcal{N}\tau_{out}}{2m} \int_{-\infty}^{\infty} dx \frac{\partial n_F(x)}{\partial x} {\rm cosh}\bigg(\frac{x}{2}\bigg)
\nonumber\\
& ~~~ \times 
\left(
\begin{array}{c}
- e^2 
\hat{Q}_s^E(x)
\\
- T x 
\hat{Q}_a^T(x)
\end{array}
\right).
\label{Sigma_x}
\end{align}
Eq.~\eqref{Sigma_x} can also be represented in the conjugate variable $\tilde{x}$
\begin{align}
\left(
\begin{array}{c}
L_{EE}
\\
L_{TT}
\end{array}
\right)
& = \frac{\pi^2 \mathcal{N}\tau_{out}}{8m} \int_{-\infty}^\infty d\tilde{x}~{\rm sech}(\pi \tilde{x})
\nonumber\\
& ~~~ \times 
\left(
\begin{array}{c}
e^2 \tilde{Q}_s^E(\tilde{x}) 
\\
- i\pi T~{\rm tanh}(\pi \tilde{x}) \tilde{Q}_a^T(\tilde{x})
\end{array}
\right).
\label{Sigma_k}
\end{align}
Plugging Eq.~\eqref{SeriesSolution_E} in Eq.~\eqref{Sigma_k}, we can write the electric conductivity 
$\sigma \equiv L_{EE}$ in the series form
\begin{align}
\frac{\sigma^{(0)}}{\sigma_0} 
& = \frac{\pi^2}{8} \frac{\tau_{out}}{\tau_{ei}} \sum_{n=0}^\infty C_{2n}^E \int_{-\infty}^\infty d\tilde{x}~{\rm sech}(\pi \tilde{x}) \tilde{Q}_{2n}(\tilde{x})
\nonumber\\
& = \frac{\tau_{out}}{\tau_{ei}} \sum_{n=0}^\infty \frac{2n + \varepsilon + 1/2}{8 [\lambda_{2n}(\varepsilon) - \Lambda_s]} \frac{\Gamma(n+1/2)\Gamma(n+\varepsilon+1/2)}{\Gamma(n+1)\Gamma(n+\varepsilon+1)}
\nonumber\\
& ~~~ \times \frac{[\Gamma(n+(\varepsilon+1)/2)]^2}{[\Gamma(n+\varepsilon/2+1)]^2},
\label{Sigma_EE_Refined}
\end{align}
in a unit of $\sigma_0 = \mathcal{N}e^2\tau_{ei} / m$.  As explained before, one can easily find the closed-form solution for $\tilde{Q}^E_s(\tilde{x})$. It follows immediately that $\sigma^{(0)} = \sigma_0$. For the eigenmode decomposition developed here, this result can be confirmed numerically; see Appendix~\ref{Appendix:ElectricConductivity}.

For the thermal conductivity, plugging Eq.~\eqref{SeriesSolution_T} in Eq.~\eqref{Sigma_k} (we identify $\kappa\approx L_{TT}$ here as explained in the main text), we find the form
\begin{align}
\frac{\kappa^{(0)}}{\kappa_0} 
& = -i \frac{3\pi}{8} \frac{\tau_{out}}{\tau_{ei}} \sum_{n=0}^\infty C_{2n+1}^T 
\nonumber\\
& ~~~ \times \int_{-\infty}^\infty d\tilde{x}~{\rm sech}(\pi \tilde{x}) {\rm tanh}(\pi \tilde{x}) \tilde{Q}_{2n+1}(\tilde{x})
\nonumber
\end{align}
in units of $\kappa_0 = \pi^2 \mathcal{N} T \tau_{ei} / (3m)$. The final result is displayed in Eq.~\eqref{Sigma_TT_Refined} of the main text.

\section{Inverse Fourier transform of $\tilde{Q}_n(\tilde{x})$}
\label{Appendix_InverseFourierTransform}

\begin{figure}[t]
\centering
\includegraphics[width=0.46\textwidth]
{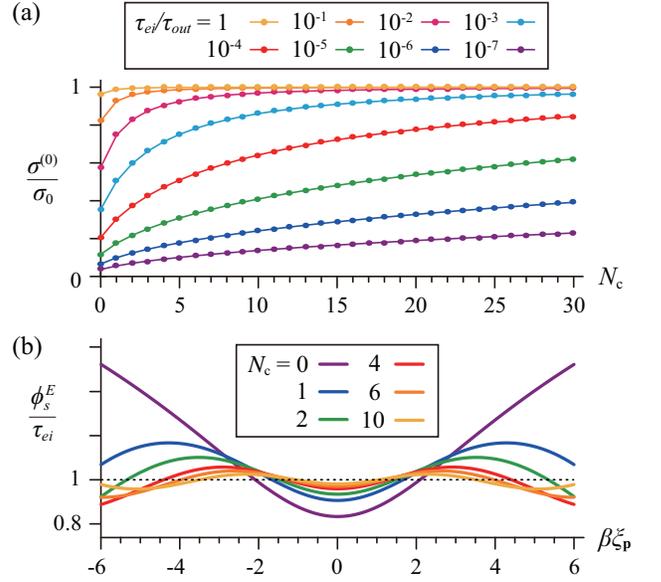} \\
\caption{
(a) Convergence of the electric conductivity $\sigma^{(0)}/\sigma_0$ [Eq.~\eqref{Sigma_EE_Refined}] as a function of the cutoff $N_c$. Here, different values of $\tau_{ei}/\tau_{out}$ are assigned to each solid line.
(b) Plot for $\phi_s^E(x)/\tau_{ei}$ as a function of $x = \beta\xi_{\bf p}$. Here, different values of $N_c$ are assigned to each solid line for fixed $\tau_{ei}/\tau_{out} = 0.5$. The black dotted line is the guideline for the solution in the RTA, [Eq.~\eqref{DistFluc_E_ConstScatt}].
}
\label{Figure9}
\end{figure}

In this section, we derive Eq.~\eqref{InverseFourierTransformQn}, the inverse Fourier transform of $\tilde{Q}_n(\tilde{x})$. It is useful to consider the following identity~\cite{Koelink96}
\begin{align}
& \int_{-\infty}^\infty d\tilde{z}~e^{-i\tilde{z}z} (1 - \tanh \tilde{z})^\alpha (1 + \tanh \tilde{z})^\beta P_n^{(\gamma,\delta)}(\tanh \tilde{z})
\nonumber\\
& = 2^{\alpha+\beta+1} \frac{\Gamma(\alpha+iz/2)\Gamma(\beta-iz/2)}{\Gamma(\alpha+\beta+n)}
\nonumber\\
& ~~~ \times i^{-n} p_n(z/2;\alpha,\delta-\beta+1,\gamma-\alpha+1,\beta),
\label{IdentityIFT}
\end{align}
which holds for ${\rm Re}(\alpha)$, ${\rm Re}(\beta) > 0$ and $- \gamma \not \in \mathbb{N}$. Here, we define the Jacobi polynomial $P_n^{(\alpha,\beta)}(\xi)$, the continuous Hahn polynomial $p_n(\zeta;a,b,c,d)$, and the Gamma function $\Gamma(z)$. The Jacobi polynomial and the continuous Hahn polynomial have the connection to the (generalized) hypergeometric function, respectively,
\begin{align}
& P_n^{(\alpha,\beta)}(\xi)
= \frac{\Gamma(n + \alpha + 1)}{\Gamma(n+1) \Gamma(\alpha + 1)}
\nonumber\\
& ~~~~~~ \times~_2F_1(-n, n + \alpha + \beta + 1, \alpha + 1, (1 - \xi)/2),
\label{JacobiPolynomial}
\end{align}
and 
\begin{align}
& p_n(\zeta;a,b,c,d) 
= i^n\frac{\Gamma(a+c+n)\Gamma(a+d+n)}{\Gamma(a+c)\Gamma(a+d)\Gamma(n+1)} 
\nonumber\\
& ~~~~ \times\!_3F_2\left(
\begin{array}{c}
-n, n+a+b+c+d-1, a+i\zeta
\\
a+c, a+d
\end{array}
; 1
\right).
\end{align}
We now evaluate the inverse Fourier transform of Eq.~\eqref{Eigenfunction_NoMomentum}
\begin{align}
\hat{Q}_n(x) 
& = \frac{\Gamma(n+1)\Gamma(\varepsilon+1)}{2\pi\Gamma(n+\varepsilon+1)} \int_{-\infty}^\infty d\tilde{x} e^{-i\tilde{x}x} [1 - {\rm tanh}(\pi\tilde{x})]^{\varepsilon/2} 
\nonumber\\
& ~~~ \times [1 + {\rm tanh}(\pi\tilde{x})]^{\varepsilon/2} P_n^{(\varepsilon,\varepsilon)}({\rm tanh}(\pi \tilde{x})),
\label{IFTQn}
\end{align}
where we used Eq.~\eqref{JacobiPolynomial} and the identity ${\rm sech}^2 (\pi\tilde{x}) = 1 - {\rm tanh}^2(\pi\tilde{x}) = [1 - {\rm tanh}(\pi\tilde{x})] [1 + {\rm tanh}(\pi\tilde{x})]$. Finally, applying Eq.~\eqref{IdentityIFT} to Eq.~\eqref{IFTQn} with replacement $2\alpha = 2\beta = \gamma = \delta = \varepsilon$, $\tilde{z} = \pi\tilde{x}$, and $z = x/\pi$, and using the identity $\Gamma(\varepsilon/2 + ix/(2\pi))\Gamma(\varepsilon/2 - ix/(2\pi)) = |\Gamma(\varepsilon/2 + ix/(2\pi))|^2$, we arrives at the resultant form, Eq.~\eqref{InverseFourierTransformQn}.

\section{Numerical evaluation of the Drude conductivity}
\label{Appendix:ElectricConductivity}

In this section, we check numerically that the eigenfunction decomposition of $\sigma^{(0)}$ reproduces the robustness of the Drude conductivity $\sigma_0$ against electron-electron scatterings in the case of a momentum-independent electron-impurity scattering rate.

In Fig.~\ref{Figure9}(a), we confirm that, whatever we choose for $\tau_{ei}/\tau_{out}$, $\sigma^{(0)}$ approaches $\sigma_0$ as increasing the cutoff $N_c$ in the summation of Eq.~\eqref{Sigma_EE_Refined}. Noticeably, the convergence is achieved with smaller $N_c$ for larger $\tau_{ei}/\tau_{out}$. A more stringent test is given by Fig.~\ref{Figure9}(b). Using Eq.~\eqref{NoneqDistFluc_Refined}, we find that $\phi_s^E(\xi_{\bf p})/\tau_{ei}$ approaches a plateau in the energy domain as increasing the cutoff $N_c$. This is fully consistent with the Keyes solution, Eq.~\eqref{DistFluc_E_ConstScatt}.

\section{Fixing $\tau_{ee}$ - Boundary condition in the clean limit}
\label{Appendix:BC}

In two and three dimensions, we commonly evaluate $\langle\!\langle \xi_{\bf p}^2 \rangle\!\rangle \approx \pi^2 T^2 / 3$ and $\langle\!\langle \xi_{\bf p} \rangle\!\rangle \approx \pi^2 T^2 / (3\epsilon_F)$ in Eq.~\eqref{ThermalConductivity_0} for $T\ll\epsilon_F$. Consistently neglecting the small quantity $\langle\!\langle \xi_{\bf p} \rangle\!\rangle^2$, we derive the thermal conductivity in the RTA in the clean limit
\begin{align}
\frac{\kappa}{\kappa_0} \approx \frac{\tau_{ee}}{\tau_{ei}}.
\label{ThermCondClean_Keyes}
\end{align}

In two dimensions, we take the same limit for the FL solution Eq.~\eqref{Sigma_TT_Refined}
\begin{align}
\frac{\kappa}{\kappa_0}
& \approx \frac{3}{8} C_2 \frac{\tau_{out}}{\tau_{ei}},
\label{ThermCondClean_2D_Refined}
\end{align}
where we define 
\begin{align}
C_2 = \sum_{n=0}^\infty 
\frac{n + 5/4}{(n+1)(n+2)(n+1/2)(n+3/2)}
= 1,
\end{align}
if we plug $\Lambda_a = 1$. Comparing Eqs.~\eqref{ThermCondClean_Keyes} and \eqref{ThermCondClean_2D_Refined}, we find the relation
\begin{align}
\frac{\tau_{ee}}{\tau_{out}}
= \frac{3}{8}.
\label{Relation_tau_ee_2D}
\end{align}

In three dimensions, the FL solution Eq.~\eqref{Sigma_TT_Refined} is approximated by
\begin{align}
\frac{\kappa}{\kappa_0}
& \approx \frac{3}{8} C_3(\Lambda_a) \frac{\tau_{out}}{\tau_{ei}},
\label{ThermCondClean_3D_Refined}
\end{align}
where we define
\begin{align}
C_3(\Lambda_a)
& = \sum_{n=0}^\infty 
\frac{n + 5/4}{(n+1)(n+3/2)[(n+1)(n+3/2) - \Lambda_a/2]}
\nonumber\\
& = \frac{1}{\Lambda_a} \bigg[ 2(1 - \ln 2) - \sum_{\gamma=\pm} H\bigg(\frac{1 + \gamma \sqrt{1 + 8\Lambda_a}}{4}\bigg) \bigg],
\label{C_3}
\end{align}
with $H(x)$ being the harmonic number for $x\in\mathbb{C}$. Comparing Eqs.~\eqref{ThermCondClean_Keyes} and \eqref{ThermCondClean_3D_Refined}, we find the relation
\begin{align}
\frac{\tau_{ee}}{\tau_{out}}
= \frac{3}{8} C_3(\Lambda_a).
\label{Relation_tau_ee_3D}
\end{align}




\begin{thebibliography}{58}
\bibitem[{Rem({\natexlab{a}})}]{Remark_phonons}
\bibinfo{note}{In this work, electron-phonon interactions are assumed to be the weakest, and they are taken into account only to connect the electrons to an external bath.}
\bibitem[{Rem({\natexlab{b}})}]{Remark_graphene}
\bibinfo{note}{A number of recent works have addressed related problems in
  transport studies of graphene, e.g. Refs.~\cite{Principi15,Xie16,Zarenia19}.
  This material has rather special transport properties, however, due to its
  linear spectrum and the peculiar role of the Dirac points.}
\bibitem[{Rem({\natexlab{c}})}]{Remark_Fermi_surface}
\bibinfo{note}{We assume that the Fermi surface is rotationally invariant. A
  thorough discussion of the generalization to non-trivial Fermi surface
  geometries can be found in Ref.~\cite{Pal12}.}
\bibitem[{rem()}]{remark_FL}
\bibinfo{note}{To align the discussions of FL and Coulomb cases, we use a
  simplified spin structure here. In a FL, it is actually more involved,
  cf.~Eqs.~(13), (14) of Ref.~\cite{Sykes70}. All results carry over to this case
  once Eq.~\eqref{eq:W} is properly adjusted.}
\bibitem[{Rem({\natexlab{a}})}]{Remark_kappa_hyd}
\bibinfo{note}{It is interesting to note that for non-degenerate electrons in
  the clean limit, the leading term in Eq.~\eqref{ThermalConductivity_0}
  reproduces the result known from classical hydrodynamics $ \kappa_{\rm hyd} =
  (1 + {d}/{2}) {\mathcal{N} T \tau_{ee}}/{m}$.}
\bibitem[{Rem({\natexlab{b}})}]{Remark_semiclassical}
\bibinfo{note}{This is in contrast to the semi-classical regime
  \cite{Andreev78,Andreev79,Spivak01} that becomes relevant at temperatures
  exceeding the Fermi energy and precedes the classical hydrodynamic regime
  \cite{Andreev11}}.
\end{thebibliography}
\end{document}